\documentclass[useAMS,usenatbib]{mnras}

\usepackage{times}
\usepackage{graphicx}
\usepackage{subfigure}
\usepackage{amsmath}
\usepackage{amssymb}
\newcommand{\rd}{{\rm d}}
\newcommand{\gsim}{\mathrel{\hbox{\rlap{\lower.55ex \hbox {$\sim$}}
                   \kern-.3em \raise.4ex \hbox{$>$}}}}
\newcommand{\lsim}{\mathrel{\hbox{\rlap{\lower.55ex \hbox {$\sim$}}
                   \kern-.3em \raise.4ex \hbox{$<$}}}}

\newcommand{\St}{\mbox{St}}

\title[Early dust coagulation during star formation]{Dust coagulation during the early stages of star formation: molecular cloud collapse and first hydrostatic core evolution}
\author[M. R. Bate]{Matthew R. Bate$^{1}$\thanks{E-mail:
M.R.Bate@exeter.ac.uk (MRB)}\\
$^{1}$ Department of Physics and Astronomy, University of Exeter, Stocker
Road, Exeter EX4 4QL  
}

\date{Accepted by MNRAS}
\begin{document}
\maketitle
\begin{abstract}
Planet formation in protoplanetary discs requires dust grains to coagulate from the sub-micron sizes that are found in the interstellar medium into much larger objects.  For the first time, we study the growth of dust grains during the earliest phases of star formation using three-dimensional hydrodynamical simulations.  We begin with a typical interstellar dust grain size distribution and study dust growth during the collapse of a molecular cloud core and the evolution of the first hydrostatic core, prior to the formation of the stellar core.  We examine how the dust size distribution evolves both spatially and temporarily.  We find that the envelope maintains its initial population of small dust grains with little growth during these phases, except that in the inner few hundreds of au the smallest grains are depleted.  However, once the first hydrostatic core forms rapid dust growth to sizes in excess of $100~\mu$m occurs within the core (before stellar core formation).  Progressively larger grains are produced at smaller distances from the centre of the core.  In rapidly-rotating molecular cloud cores, the `first hydrostatic core' that forms is better described as a pre-stellar disc that may be gravitationally unstable.  In such cases, grain growth is more rapid in the spiral density waves leading to the larger grains being preferentially found in the spiral waves even though there is no migration of grains relative to the gas.  Thus, the grain size distribution can vary substantially in the first core/pre-stellar disc even at these very early times.
\end{abstract}

\section{Introduction}
\label{introduction}

Dust plays important roles in both star formation and planet formation.  In the former case, dust thermal emission dominates molecular gas cooling at number densities $n \gsim 10^4$~cm$^{-3}$ \citep[e.g.,][]{Goldsmith2001}.  In the later case, the growth of interstellar grains in the comparatively high-density environments provided by protoplanetary discs eventually leads to planet formation.

Dust in the interstellar medium is typically believed to have a maximum grain size of $s \sim 0.1~\mu$m  and frequently modelled with an MRN size distribution \citep*[][]{MatRumNor1977}.   The coagulation time scale for grains at these low densities is very long, and larger grains are thought to be destroyed in interstellar shocks.
Within protoplanetary discs, however, the much higher densities and their more quiescent nature allow substantial growth of dust grains  \citep[e.g.,][]{WeiCuz1993}.  The growth and dynamical evolution of dust grains in protoplanetary discs has been the subject of many studies, both theoretical and observational, over many years.

Much less attention has been paid to dust growth during the earlier star formation and protoplanetary disc formation phases, although a number of studies have considered this phase.   \cite{Ormel_etal2009} considered whether or not large dust grains could grow within molecular cloud cores.  They found that large grains could grow within molecular clouds, but only if they were both dense and long-lived.
\cite{HirOmu2009} modelled dust growth at the centre of a collapsing molecular cloud core, with metallicities ranging from solar ($Z={\rm Z}_\odot$) to $Z=10^{-6} ~{\rm Z}_\odot$.  Beginning with a typical MRN  size distribution, they showed that there was little dust coagulation during the early phase of the collapse (densities $n_{\rm H} \lesssim 10^7$~cm$^{-3}$ in any of their calculations), but that as the density increased substantial grain growth may occur for high metallicities.  For metallicities above $\sim 10^{-4} ~{\rm Z}_\odot$ the smallest grains became depleted at the centre of the first hydrostatic core ($n_{\rm H} \gtrsim 10^{11}$~cm$^{-3}$), but only for solar metallicities did the peak grain size (in terms of dust mass) increase beyond its initial value of 0.25~$\mu$m, and it remained sub-micron.  Recently, \cite{Guillet_etal2020} also used one-dimensional collapse models to examine the effects of dust coagulation on magnetic resistivity during the collapse of molecular cloud cores.  They found that hydrodynamic turbulence is inefficient at removing small grains, but that additional drift velocities induced by ambipolar diffusion may enhance dust coagulation.

Recent observations of Class 0 and I protostellar objects find that their envelopes apparently contain mm-size dust grains (\citealt{Kwonetal2009}; \citealt*{ChiLooTob2012}; \citealt{Miotello_etal2014,Li_etal2017,Galametz_etal2019}), despite their young ages ($\sim 10^5$ yrs).   \cite*{WonHirLi2016}  investigated whether or not mm-size dust grains could grow at the densities found within these envelopes and concluded that they could not.  Instead, they proposed that the large grains were produced in the higher-density discs of Class 0 and I protostellar objects and launched into the envelope via protostellar outflows.  \cite{BatLor2017} considered the possibility that molecular clouds already contain some large dust grains and modelled their dynamics during the hydrodynamic collapse of a molecular cloud.  They showed that if dust with sizes $s \gsim 100 ~\mu$m is present, it evolves quite differently from the gas and its distribution may be quite dissimilar from that of the gas even before stellar core formation occurs. However, dust grains with sizes $s \lsim 10 ~\mu$m closely follow the collapsing molecular gas.  Similar results have been reported by \cite*{LebComLai2020} who performed both hydrodynamical and magnetohydrodynamical simulations.

In this paper, we implement the dust grain growth scheme of \cite{HirOmu2009} into a three-dimensional smoothed particle hydrodynamics (SPH) code to investigate dust growth during the early stages of star formation.  We study the evolution up to just prior to stellar core formation.  This can be viewed as a natural extension of \cite{HirOmu2009} who considered a `single zone' model of dust growth during the collapse of a molecular cloud core (i.e., they followed dust growth at the densest point at the centre of a collapsing cloud, but couldn't examine the spatial variation).  By implementing dust growth for each SPH particle in a three-dimensional simulation, we are able to investigate both the temporal and spatial distributions of dust particles sizes during the molecular cloud collapse and the formation and evolution of the first hydrostatic core \citep{Larson1969} or pre-stellar disc \citep[rapidly-rotating first hydrostatic cores are disc-like in morphology and can have radii of several tens of au;][]{Bate1998, Bate2011}.

However, in this first paper, we only consider dust growth.  We do not treat differential migration of dust particles, either relative to the gas, or of different sizes relative to each other.  In other words, the amount of dust associated with each SPH particle evolves independently of nearby dust that is modelled by other SPH particles.  This approximation is valid until dust grains grow large enough to move relative to each other, or relative to the gas, over length-scales that are resolved by the simulation. Beyond that, dust dynamics needs to be taken into account.  We demonstrate, however, that this does not become a limiting issue in any of the simulations presented here because if the dust initially has a typical MRN size distribution it does not grow sufficiently large over the timescales that we model.  Furthermore, because the dust grain distribution modelled by each SPH particle is independent of every other SPH particle, parallelisation of this part of the code is trivial.  

This paper is arranged as follows.  In Section \ref{sec:method} we describe dust growth method and its implementation into the SPH code.  We also briefly describe the SPH code itself (Section \ref{sec:sph}), and the initial conditions for the calculations (Section \ref{sec:initialconditions}).  In Section \ref{sec:results}, we present results from star formation calculations, beginning with spherical collapse (a non-rotating molecular cloud core; Section \ref{sec:spherical}), and then rotating collapse calculations (Section \ref{sec:rotating}).  We examine both the temporal and spatial evolution of the dust size distribution, and we consider the limitations of the calculations (Section \ref{sec:caveats}).  Finally, in Section \ref{sec:conclusions} we present our conclusions.

\section{Method}
\label{sec:method}

First, we discuss our model for grain growth and its limitations.  Second, we briefly describe the smoothed particle hydrodynamics (SPH) code in which the grain growth method is implemented.  Finally, we describe the initial conditions for our SPH calculations.

\subsection{Grain growth model}

We use a grain growth method very similar to that of \cite{HirYan2009}, that was also used by  \cite{HirOmu2009} to study dust coagulation at the centre of collapsing of molecular clouds of various metallicities.  They developed the method to model grain growth in a single zone using $N$ discrete bins to model the distribution of grains with different sizes.  We essentially use the same method, but with the grain population of each individual SPH gas particle modelled as a separate `zone'.  Thus, we treat the grain growth as occurring independently of neighbouring zones (i.e. neighbouring SPH particles).  This is a good approximation as long as the real movement of grains between zones is negligible or neighbouring zones have a similar grain population.  This limitation of treating each SPH particle independently is discussed further in Section \ref{sec:limitations}.

In this section we summarise the grain growth model of \cite{HirYan2009} and \cite{HirOmu2009}; the interested reader is directed to those two papers for more detail.
We assume spherical grains with a constant material density $\rho_{\rm gr}$ such that the mass, $m$, of a grain of radius, $a$, is given by
\begin{equation}
m(a)= \frac{4 \pi}{3} a^3 \rho_{\rm gr}.
\end{equation}
The number density of grains with radii between $a$ and $a+\rd a$ is given by $n(a)~\rd a$ and we model grains with sizes ranging from $a_{\rm min}$ to $a_{\rm max}$.  To ensure conservation of total grain mass \cite{HirYan2009} consider the distribution function of grain mass rather than grain size, denoting the number density of grains with masses between $m$ and $m+{\rm d}m$ as $\tilde{n}(m)~{\rm d}m$, where the two distributions are related as $n(a)~\rd a = \tilde{n}(m)~\rd m$.  Note that because of the way these are defined $n(a)$ alone actually has units of cm$^{-4}$ and $\tilde{n}(m)$ alone actually has units of cm$^{-4}$g$^{-1}$, so that physically it only really makes sense to think of $n(a)~\rd a$ or $\tilde{n}(m)~\rd m$, which give the number density of grains in some size or mass interval, respectively.

For each SPH particle, we consider $N$ discrete bins that are logarithmically spaced in grain radius \citep[see ][for further detail]{HirYan2009}.  The grain radius in the $i$th bin lies between $a^{(b)}_{i-1}$ and $a^{(b)}_i$ where $a^{(b)}_i = \delta a^{(b)}_{i-1}$ and $\log(\delta) = \log(a_{\max}/a_{\min})/N$.  The grain radius and the mass in the $i$th bin are defined as $a_i \equiv (a^{(b)}_{i-1} + a^{(b)}_{i})/2$ and $m_i \equiv m(a_i)$, respectively.  Note that the grain radius could be defined using a geometric mean rather than an arithmetic mean.  We have tried both and find no significant difference, at least for the number of bins that we use, so we choose to follow \cite{HirYan2009}. The upper boundary of the mass bin is defined as $m^{(b)}_i \equiv m(a^{(b)}_i)$.

The mass density of grains in the $i$th bin, $\tilde{\rho}_i$, is defined as
\begin{equation}
\tilde{\rho}_i \equiv m_i \tilde{n}(m_i) \left[ m^{(b)}_i - m^{(b)}_{i-1} \right].
\end{equation} 
The time evolution of $\tilde{\rho}_i$ can then be expressed as
\begin{equation}
\frac{ \rd \tilde{\rho}_i }{\rd t} = \left[ \frac{ \rd \tilde{\rho}_i }{\rd t} \right]_{\rm coag}, 
\end{equation}
where we only consider coagulation.  \cite{HirYan2009} considered both coagulation and shattering because they were interested in studying grain evolution in interstellar turbulence, while \cite{HirOmu2009} only considered coagulation because they were concerned with modelling early grain evolution in the relatively quiescent environment of a molecular cloud core in which shattering is negligible.  In this paper, we are also only concerned with early grain growth in a molecular cloud core or early stage protostellar disc containing relatively small grains that are quite well coupled to the gas, so for this paper we also only include coagulation.   Furthermore, for simplicity, we only consider a single grain species with $\rho_{\rm gr}=2.26$~g~cm$^{-3}$ \citep[][considered a mixture of silicate $\rho_{\rm gr}=3.3$~g~cm$^{-3}$ and graphite $\rho_{\rm gr}=2.2$~g~cm$^{-3}$ grains, although grains of different types couldn't coagulate]{HirOmu2009}.

The time evolution of $\tilde{\rho}_i$ due to coagulation can be written as
\begin{equation}
\left[ \frac{ \rd \tilde{\rho}_i }{\rd t} \right]_{\rm coag} = - m_i \tilde{\rho}_i \sum^N_{k=1} \alpha_{ki} \tilde{\rho}_k + \frac{1}{2} \sum^N_{j=1} \sum^N_{k=1} \alpha_{kj} \tilde{\rho}_k \tilde{\rho}_j m^{kj}_{\rm coag}(i),
\label{eq:coag}
\end{equation}
where
\begin{equation}
 \alpha_{kj} = \left\{  \begin{array}{ll} 
\displaystyle \frac{\beta \sigma_{kj} v_{kj}}{m_j m_k}  & {\rm if}~~ v_{kj} < v^{kj}_{\rm coag}, \\ 
\\
 0 & {\rm otherwise.} 
 \end{array} \right.
 \label{eq:alpha}
\end{equation}
In equation \ref{eq:coag}, \cite{HirYan2009} and \cite{HirOmu2009} use $m^{kj}_{\rm coag}(i) = m_i$ if the mass of the coagulated particle $m_k + m_j$ is within the mass range of the $i$th bin, otherwise $m^{kj}_{\rm coag}(i) = 0$.  However, we found that this does not result in good dust mass conservation.  Instead, we use $m^{kj}_{\rm coag}(i) = m_k+m_j$ if this mass is within the mass range of the $i$th bin, which not only conserves total dust mass better, but also results in a more accurate solution with small numbers of bins.  Note also that the factor of $1/2$ that appears in equation \ref{eq:coag} is missing in the equivalent equation in \cite{HirYan2009}, but corrected in \cite{HirOmu2009}.  

Coagulation only occurs if the relative velocity, $v_{ki}=| \mbox{\boldmath{$v$}}_k - \mbox{\boldmath{$v$}}_i|$, is less than the coagulation threshold velocity $v^{ki}_{\rm coag}$ (see equation \ref{eq:alpha}).  The cross section for coagulation is $\sigma_{ki} = \pi (a_k+a_i)^2$ and $\beta$ is the sticking probability which we set to unity \citep[since][show that the sticking probability is of the order of unity below the coagulation threshold velocity]{Blum2000}.  We use the coagulation threshold velocity \citep*{ChoTieHol1993,DomTie1997,YanLazDra2004,HirOmu2009}
\begin{equation}
v^{ki}_{\rm coag} = 21.4 \left[ \frac{a_k^3 + a_i^3}{(a_k+a_i)^3} \right]^{1/2} \frac{\gamma_{\rm SE}^{5/6}}{E^{1/3} R_{ki}^{5/6} \rho_{\rm gr}^{1/2}},
\end{equation}
where $\gamma_{\rm SE}$ is the surface energy per unit area, $R_{ki} \equiv a_k a_i/(a_k+a_i)$ is the reduced radius of the grains, and $E$ is related to Poisson's ratios ($\nu_k$ and $\nu_i$) and Young's modulus ($E_k$ and $E_i$) by $1/E \equiv (1-\nu_k)^2/E_{k} + (1-\nu_i)^2/E_{i}$.  We use the same values for these parameters as \cite{HirYan2009} use for graphite grains: $\gamma_{\rm SE}=12$~erg~cm$^{-2}$, $E=3.4\times 10^{10}$~dyn~cm$^{-2}$.  For two identical particles, these parameters result in $v^{ii}_{\rm coag} \approx 228~ (a_i/\mbox{0.1 $\mu$m})^{-5/6}$~cm~s$^{-1}$.

\subsection{Implementation}
\label{sec:implementation}

In terms of the implementation of the above method in the SPH code, each Lagrangian SPH particle is treated as a single `zone' model that is evolved independently of every other `zone'.  Furthermore, rather than evolve $\tilde{\rho}_i$ directly, we set $\tilde{\rho}_i = \varrho_i  \rho$, where $\rho$ is the total SPH density (gas and dust) of the particle, and evolve $\varrho_i$.  In other words, we evolve the mass density of grains in the $i$th bin as a fraction of the total mass density.  This allows us to naturally include the compression of the Lagrangian `zone' during collapse (i.e., as the cloud collapses the density of grains increases simply due to the compression of the `zone').

A typical interstellar grain size distribution is often specified as a power-law distribution for sizes smaller than some maximum, cut-off size (e.g., $N(a) \propto a^{-3.5}$ for $a<a_{\rm cutoff}$).  Therefore, initially we set $\varrho_i$ using the chosen dust-to-gas ratio, $\varepsilon$, and the dust grain size distribution, $N(a)$  such that
\begin{equation}
\varrho_i = \frac{ \varepsilon }{\chi}  \displaystyle  \int_{a^{(b)}_{i-1}}^{a^{(b)}_{i}} a^3 N(a)~ {\rm d}a ,
\end{equation}
where the normalisation factor is
\begin{equation}
\chi = \displaystyle \int_{a_{\rm min}}^{a_{\rm cutoff}} a^3 N(a)~ {\rm d}a,
\end{equation}
since the dust grain mass with radii between $a$ and $a+{\rm d}a$ is 
\begin{equation}
m(a) N(a) {\rm d}a = \frac{4 \pi \rho_{\rm gr}}{3} a^3 N(a)~ {\rm d}a = \frac{4 \pi \rho_{\rm gr}}{3} a^4 N(a)~ {\rm d}(\log a).
\label{eq:mass_dist}
\end{equation}
This means that $\sum_i \varrho_i = \varepsilon$ for each `zone' (which should be constant since `zones' do not transfer grains).  Furthermore, in the results section of this paper, we plot the grain mass distribution in a single `zone' as
\begin{equation}
\frac{ \varrho_i }{\varepsilon}  \frac{1}{ \log\delta} ~~~~ = ~~~~ 4 \pi \rho_{\rm gr}/3 ~ a^4 N_{\rm gr}(a),
\label{eq:mass_dist}
\end{equation}
which differs in normalisation by the factor $4 \pi \rho_{\rm gr}/3$ from the quantity $a^4 N_{\rm gr}(a)$ that is plotted by \cite{HirOmu2009} (e.g., their figure 3).

\subsection{Limitations}
\label{sec:limitations}

In this paper, we consider grain growth, but we do not explicitly allow grains to migrate relative to the gas.  Thus, our method is only applicable to small dust grains that are well coupled to the gas.  When does this approximation begin to break down?

It is generally assumed that dust in the interstellar medium (ISM) consists of small grains ($\sim 0.1 ~\mu$m) so that in molecular clouds the dust is both well-mixed with, and well-coupled to, the gas (but see \cite{Hopkins2014,HopLee2016} who argue that in the low-density ISM even small grains can become poorly coupled and, therefore, some dust-to-gas ratio variation may be expected, and \cite*{TriPriLai2017} who simulate the dynamics of dust grains in turbulent molecular clouds).  

The dynamical interaction between gas and dust is manifest primarily as a drag force.  The drag force experienced by dust grains moving through a gas depends mainly on the size of the dust grains and the density of the gas \citep{Whipple1972, Weidenschilling1977}.  If the mean free path of the gas molecules is larger than the dust particle's radius, the characteristic timescale for the decay of the dust particle's speed relative to the gas, the stopping time, can be expressed as
\begin{equation}
t_{\rm s} = \frac{\rho_{\rm gr} s}{\rho_{\rm G} v_{\rm th} },
\label{eq:stoppingtime}
\end{equation}
where, for simplicity, we have assumed that the dust-to-gas ratio is small, $s$ is the radius of the dust particle, $\rho_{\rm G}$ is the gas density, and the velocity of the gas molecules due to thermal motion is
\begin{equation}
v_{\rm th} = \sqrt{\frac{8 k_{\rm B} T}{\pi \mu m_{\rm H}}},
\end{equation}
where $T$ is the gas temperature, $\mu$ is the mean molecular weight of the gas, $m_{\rm H}$ is the atomic mass of hydrogen, and $k_{\rm B}$ is Boltzmann's constant. Thus, small grains have short stopping times and are better coupled to the gas dynamically, while larger grains have longer stopping times and may be only weakly affected by the drag force.

For a mixture of gas and dust, differential speeds are driven by the fact that the gas experiences a net pressure force, while the dust does not.  In the terminal velocity limit, this leads to a differential speed
\begin{equation}
v_{\rm DG} \rightarrow t_{\rm s} \nabla P_{\rm G}/\rho_{\rm G},
\label{eq:terminalv}
\end{equation}
assuming a low dust-to-gas ratio.
Using this terminal velocity approximation, we can estimate when the differential speed between the gas and dust will become too large for our assumption that the grains remain local to be valid.  We can express the gas pressure as 
\begin{equation}
P_{\rm G} = \frac{k_{\rm B} T}{\mu m_{\rm H}}  \rho_{\rm G}.  
\label{eq:gasP}
\end{equation}
We can also assume that the maximum density gradient that can be resolved in an SPH simulation is $\nabla \rho_{\rm G}\sim \rho_{\rm G}/h$, where $h$ is the SPH particle smoothing length, so that $(\nabla \rho_{\rm G})/\rho_{\rm G}\sim 1/h$.  Combining these equations, we can write
\begin{equation}
v_{\rm DG} \approx \frac{\rho_{\rm gr} s}{\rho_{\rm G} h}\sqrt{\frac{\pi k_{\rm B}T}{8 \mu m_{\rm H}}}.
\label{eq:vDG}
\end{equation}
The time required for the dust particle to move the SPH resolution length, $h$, at its terminal velocity is 
\begin{equation}
t_{\rm valid} \lesssim \frac{h}{v_{\rm DG}} = \frac{\rho_{\rm G} h^2}{\rho_{\rm gr} s}\sqrt{\frac{8 \mu m_{\rm H}}{\pi k_{\rm B}T}}.
\label{eq:tvalid1}
\end{equation}
We can further assume that for an SPH simulation with particles of equal-mass, $m_{\rm p}$, that $h \approx (m_{\rm p}/\rho_{\rm G})^{1/3}$ so that
\begin{equation}
t_{\rm valid} \lesssim \frac{(\rho_{\rm G} m_{\rm p}^2)^{1/3}}{\rho_{\rm gr} s}\sqrt{\frac{8 \mu m_{\rm H}}{\pi k_{\rm B}T}}.
\label{eq:tvalid2}
\end{equation}
For example, for a dust grain of density $\rho_{\rm gr}=2.26$~g~cm$^{-3}$ and size 0.1~$\mu$m in a 1-M$_\odot$ gas cloud of number density $10^5$~cm$^{-3}$ at $T=10$~K that is modelled by $10^6$ SPH particles, this time scale is $t_{\rm valid} \lesssim 1.4 \times 10^5$ yr, which is similar to the free-fall time.  This is long enough to model the evolution of a collapsing dense molecular cloud core, and the timescale increases as the density increases (as long as the dust does not grow quickly, and the temperature does not increase quickly).  Note also that if there is a size distribution of dust grains associated with each SPH gas particle and $s$ is taken to be the size of the largest dust grains, then the smaller dust grains will satisfy this criterion more easily. 

\begin{figure}
\centering \vspace{-0.25cm} \vspace{0cm}
    \includegraphics[width=8.8cm]{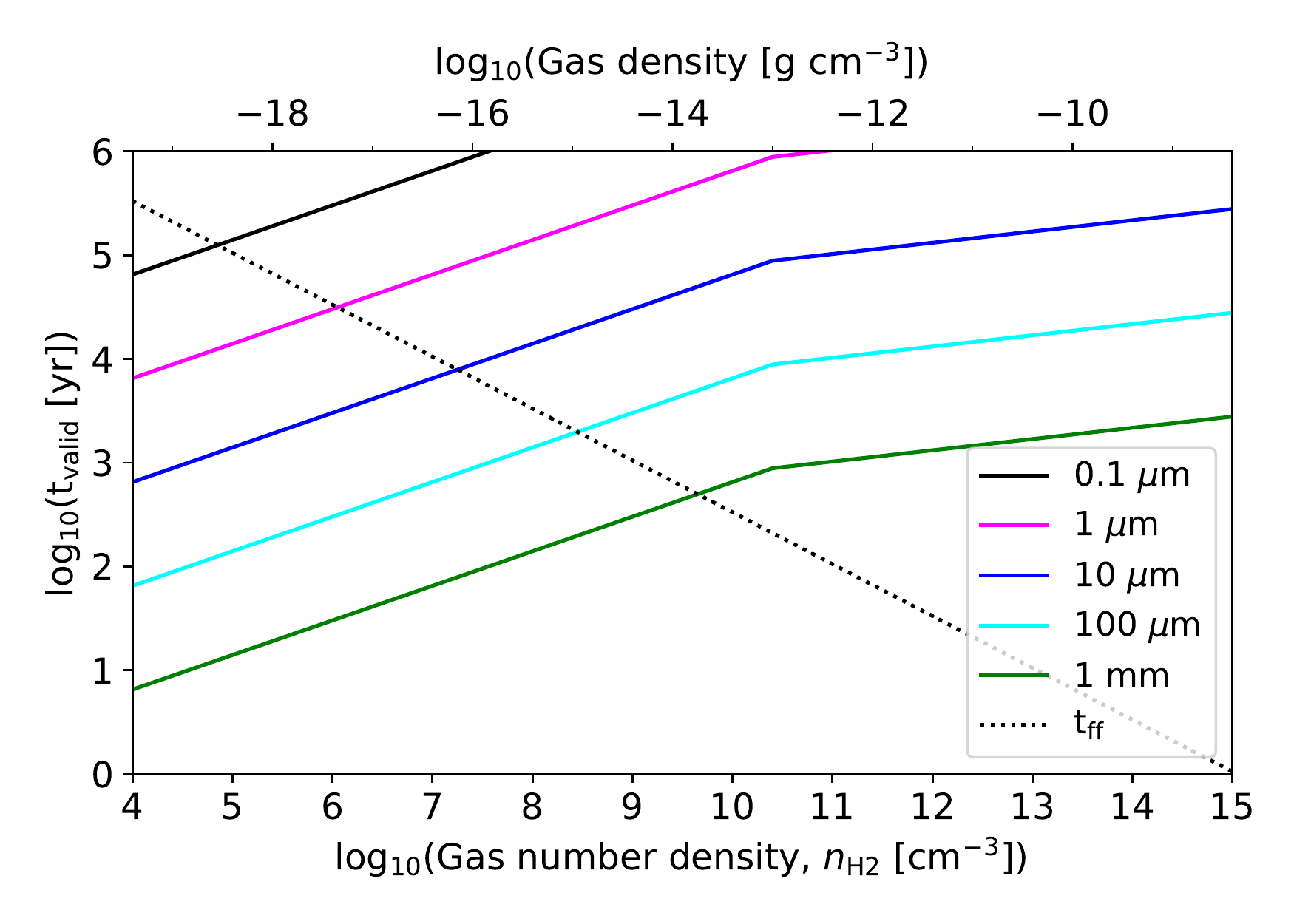} \vspace{-0.5cm}
\caption{The timescale for dust grains with different sizes moving relative to the gas at their terminal velocity to move the local resolution length, $h$, in an SPH simulation of the collapse of a 1-M$_\odot$ molecular cloud core modelled using $10^6$ SPH particles  (equation \ref{eq:tvalid2}).  The plot assumes spherical dust grains of intrinsic density $\rho_{\rm gr}=2.26$~g~cm$^{-3}$, a gas mean molecular weight of $\mu=2.38$ and a gas temperature of $T=10$~K for densities $\rho_{\rm G} \leq 10^{-13}$~g~cm$^{-3}$ which increases at higher densities as $\log{T}=1+ (\log{\rho_{\rm G}}+13)^{0.45}$ which mimics the typical evolution of the temperature at the centre of a collapsing molecular cloud core, such as that seen in Fig.~\ref{fig:dens_temp}.  The free-fall timescale is given by the dotted black line for comparison.  As long as the grains are small enough and/or the cloud density is high enough (upper left region), it is valid to neglect the diffusion of dust grains within the simulation.}
\label{fig:tvalid}
\end{figure}

In Fig.~\ref{fig:tvalid}, we plot this validity timescale as a function of gas density and of the size of the largest dust grain.  The plot assumes that a 1-M$_\odot$ cloud is modelled using $10^6$ SPH particles.  It uses an approximation to the gas temperature (see the figure caption) that gives a reasonable approximation throughout the first hydrostatic core phase phase and up to $\rho_{\rm G} \approx 10^{-8}$~g~cm$^{-3}$ and $T \approx 2000$~K, whereupon molecular hydrogen dissociation triggers the second collapse phase (and dust grains would evaporate).  As long as the timescale, $t_{\rm valid}$, is longer than the time of the simulation (or the time at which this timescale has been relevant for), then our approximation that the dust grains remain local (i.e., within an SPH gas particle smoothing length) should be valid.  In practice, when we perform the simulations for this paper we monitor how far the typical dust grain should have moved as a fraction of the SPH particle smoothing length to confirm that this approximation remains valid (see Appendix \ref{appendixB}).

\subsection{Dust grain relative velocities}

The above timescale for validity of the simulations assumes that the velocity difference between the gas and the dust is due simply to a gas pressure gradient.  It also doesn't consider relative motions between dust particles (e.g., with different sizes).  In reality there are several sources of dust particle motion.  Furthermore, dust grain growth depends on dust particles having different velocities so that they can collide and coagulate ($v_{ki}$ in equation \ref{eq:alpha}).  

The relative motions of the smallest dust particles are dominated by Brownian motion due to their bombardment by gas particles.  If the gas is turbulent, gas drag also provides a source of dust relative velocity.  Finally, in a protoplanetary disc, gas drag in the equilibrium disc structure leads to radial and azimuthal drift of dust particles, and vertical settling toward the disc's midplane, all of which vary with the size of the dust particle and, thus, provide dust relative velocities.  In this paper, we investigate the effects of most of these sources of relative velocity.

\subsubsection{Brownian motion}
\label{sec:Brownian}

\cite{HirOmu2009} adopted a single velocity for each grain size based on the thermal (Brownian) velocity
\begin{equation}
v_k = \sqrt{\frac{8 k_{\rm B}T}{\pi m_k}}.
\label{eq:brownian}
\end{equation}
Smaller grains have larger thermal velocities so the effect on coagulation is strongest for the smallest grains (see Section \ref{sec:results}).  \cite{HirOmu2009} divided each timestep into four equal small steps and applied a different relative grain velocity in each sub-step: $v_{ki}= v_k + v_i$, $|v_k - v_i|$, $v_k$ and $v_i$.  Another common method of setting the relative velocity in the Brownian motion regime is to use 
\begin{equation}
\left[ v_{ki} \right]_{\rm Brownian}= \sqrt{\frac{8 k_{\rm B}T}{\pi \mu_{ki}}}.
\label{eq:brownian_reduced}
\end{equation}
where $\mu_{ki}$ is the reduced mass of the two grains \citep*[e.g.,][]{BirFanJoh2016}.  We use this latter method for setting the relative velocities due to Brownian motion.

\subsubsection{Terminal velocities during gravitational collapse}
\label{sec:pres_grad}

Fundamentally, many sources of relative motion are due to the pressure acceleration on the dust being negligible (because $\rho_{\rm gr} \ll \rho_{\rm G}$).  Thus, the gas experiences an extra acceleration compared to the dust.  In the purely radial collapse of a molecular cloud core (i.e., neglecting rotation), this leads to the gas being partially supported against gravity and, therefore, collapsing more slowly than dust particles.  Furthermore, as the dust collapses it experiences a drag force from the gas which depends on the dust particle's size and, therefore, leads to relative dust velocities.  Large grains collapse more quickly than the gas or smaller dust particles \citep[e.g.,][]{BatLor2017}.

Assuming purely radial infall and small dust grains for which the terminal velocity approximation applies (i.e., whose stopping times are much shorter than the free-fall time), the relative velocity between two dust particles can be expressed as
\begin{equation}
\left[ v_{ki} \right]_{\rm pres} \approx  (t_{\rm s,k} - t_{\rm s,i}) \nabla P_{\rm G}/\rho_{\rm G}.
\label{eq:terminalv}
\end{equation}
We apply this source of relative dust velocities for SPH particles that are considered to be part of the infalling gas envelope (see below).

\subsubsection{Turbulence}
\label{sec:turbulence}

Another source of dust relative motions that we consider in this paper is due to gas turbulence, which the dust grains are coupled to via drag \citep{Ossenkopf1993, WeiRuz1994, Ormel_etal2009}.  We investigate the effects of turbulence in the collapsing envelope as well as the turbulence within a disc.

When considering dust dynamics in protoplanetary discs it is common to use the Stokes number, ${\rm St}= t_{\rm s}\Omega$, where  $\Omega$ is the orbital angular frequency.  We use the parameterisation of \cite{OrmCuz2007} to treat dust relative motions due to turbulent motions in a disc.  \citeauthor{OrmCuz2007} consider both class I (slow) eddies and class II (fast) eddies.  We are only concerned with relatively small dust particles that are relatively well coupled to the gas.  Therefore, we implement equations 26 and 28 from \cite{OrmCuz2007} which cover the `tightly coupled' (class I eddies) and the `fully intermediate' (class II eddies) regimes.  For the former:
\begin{equation}
\left[ v_{12} \right]^2_{\rm turb,I} = V_{\rm g}^2  \frac{\St_1 - \St_2}{\St_1 + \St_2} \left( \frac{\St_1^2}{\St_{1} + {\rm Re}^{-1/2}}  - \frac{\St_2^2}{\St_{2} + {\rm Re}^{-1/2}} \right) 
\label{eq:class1}
\end{equation}
where the two dust grains are ordered such that ${\rm St}_1 \geq {\rm St}_2$, and Re is the Reynolds number.
For the latter:
\begin{equation}
\left[ v_{12} \right]^2_{\rm turb,II} = V_{\rm g}^2 {\rm St}_1 \left[ 2 y_{\rm a} - (1+\epsilon) + \frac{2}{1+\epsilon} \left( \frac{1}{1+y_{\rm a}} + \frac{\epsilon^3}{y_{\rm a} + \epsilon} \right) \right],
\end{equation}
where $\epsilon=\St_2/\St_1 \leq 1$ and we take $y_{\rm a}=1.6$.  This latter contribution from class II eddies is only applied when  $\St_1>{\rm Re}^{-1/2}/y_{\rm a}$.

In the above equations, we limit St$_2$ to be no greater than $0.9~{\rm St}_1$ so that even dust particles of the same size are assumed to have some relative motion due to the turbulence (otherwise equation \ref{eq:class1} gives zero relative velocity for two identical dust grains).  Such forcing of identical dust grains to have some relative motion is commonly done in dust growth studies (e.g., \citealt{Okuzumi_etal2012,Krijt_etal2016}; \citealt*{SatOkuIda2016}) and is justified by the fact that, in reality, two dust grains of equal mass will not have identical shapes.

For the turbulence in the disc, we assume an $\alpha_{\rm SS}$-disc model \citep{ShaSun1973}, the normalisation of the contributions to the relative grain velocities from turbulence in a disc can be expressed as $V_{\rm g}=\alpha_{\rm SS}^{1/2} c_{\rm s}$ \citep*[e.g.,][]{NakSekHay1986,CuzWei2006}, and the Reynolds number as ${\rm Re} = \alpha_{\rm SS} c_{\rm s} H/\nu = \alpha_{\rm SS} c_{\rm s}^2/(\nu \Omega)$ where $c_{\rm s}=v_{\rm th} \sqrt{\pi/8} $ is the isothermal gas sound speed, $H$ is the gas disc scale height, and $\Omega$ is the orbital angular frequency.  For the calculations reported in this paper, we set $\alpha_{\rm SS}=10^{-3}$ (to set $V_{\rm g}$), but for simplicity, we simply set the Reynolds number to a constant value of ${\rm Re} = 10^8$.  

For the turbulence in the envelope, we assume $V_{\rm g}=(3/2)^{1/2} c_{\rm s}$ (e.g., \citealt{Ormel_etal2009,Guillet_etal2020}), and we define the Stokes number in the envelope as ${\rm St} = t_{\rm s} c_{\rm s}/\lambda_{\rm Jeans} = 2 t_{\rm s} \sqrt{G \rho_{\rm G}/\pi}$ (e.g., \citealt{Guillet_etal2020}), where the Jeans length $\lambda_{\rm Jeans}$ has been defined by relating the sound crossing time to the free-fall time.  Again, for simplicity we set the Reynolds number to a constant value of ${\rm Re} = 10^8$.

\subsubsection{Pressure gradients in a disc}
\label{sec:pres_disc}

When gas is orbiting a central object (e.g., a star), the acceleration due to pressure gradients that is experienced by the gas leads to different effects such as gravitational settling of dust toward the disc mid-plane, radial dust drift, and azimuthal velocity differences between the gas and dust particles \citep{Whipple1972,Weidenschilling1977,NakSekHay1986}.  The relative grain velocities produced by all of these processes depend on the dust grain size, again providing a source of relative motion between dust grains.

We only implement grain relative velocities due to vertical settling and radial drift motions because for the Stokes numbers we are interested in these relative velocities are much larger than the relative azimuthal velocities \citep[see, e.g.,][]{BirFanJoh2016}.  The usual parameterisation of the relative grain velocities due to radial drift in a disc is
\begin{equation}
\left[ v_{ki} \right]_{\rm radial} = -2~ |{\rm St}_k - {\rm St}_i| ~ \eta ~v_{\rm Kep},
\label{eq:rad_drift1}
\end{equation}
where $\eta = - \frac{1}{2} (H/r)^2 ( \partial\ln P / \partial\ln r)$ measures how much slower the gas is orbiting compared to the Keplerian speed (i.e., $v_{\phi,G} = \eta ~v_{\rm Kep}$).  Rather than evaluating $\eta$ in this form, however, we instead calculate $\eta = - \frac{1}{2} ( \mbox{\boldmath{$a$}}_{\rm gas} \cdot \mbox{\boldmath{$a$}}_{\rm grav})/a_{\rm grav}^2$, where $\mbox{\boldmath{$a$}}_{\rm gas}$ is the acceleration on an SPH particle due to pressure only, and $\mbox{\boldmath{$a$}}_{\rm grav}$ is the acceleration on the SPH particle due to gravity only.  We also avoid having to compute $v_{\rm Kep}$ directly by in fact using
\begin{equation}
\left[ v_{ki} \right]_{\rm radial} = -2~ | t_{{\rm s},k} -  t_{{\rm s},i}  | ~ \eta ~a_{\rm grav},
\label{eq:rad_drift2}
\end{equation}
since $a_{\rm grav} = \Omega v_{\rm Kep}$.

For the relative velocities of small grains due to dust settling in a disc we use the usual expression that assumes that the grains are settling at their terminal velocity \citep[e.g.,][]{BirFanJoh2016}:
\begin{equation}
\left[ v_{ki} \right]_{\rm settle} = z\Omega | {\rm St}_k - {\rm St}_i |.
\label{eq:settling}
\end{equation}

\subsubsection{Combining the different sources of relative dust velocities}

Each of the above sources of relative dust velocity are computed separately and combined in quadrature to give the overall relative grain velocity that appears in equation \ref{eq:alpha}, i.e.,
\begin{equation}
\begin{split}
v_{ki}^2 =  \left[ v_{ki} \right]_{\rm Brownian}^2 + & \left[ v_{ki} \right]_{\rm pres}^2 + \left[ v_{ki} \right]_{\rm turb,I}^2 +  \left[ v_{ki} \right]_{\rm turb,II}^2 + \\
&  \left[ v_{ki} \right]_{\rm radial}^2 +  \left[ v_{ki} \right]_{\rm settle}^2.
\end{split}
\label{eq:quadrature}
\end{equation}
However, the $\left[ v_{ki} \right]_{\rm pres}$ term is only applicable in the collapsing envelope of the molecular cloud core, while the latter two terms are only applicable in a disc, and the functional form of the two turbulent terms varies between the envelope and the disc.  Therefore, we need some way to determine whether an SPH particle is an `envelope' or a `disc' particle.

We define an SPH particle to lie in a disc if the magnitude of its radial velocity is less than half the magnitude of its tangential velocity.  We computed these velocities relative to the direction of the gravitational acceleration experienced by the SPH particle.  In equation form, the radial velocity is $v_{\rm rad} = \mbox{\boldmath{$v$}} \cdot \hat{\mbox{\boldmath{$a$}}}_{\rm grav}$ and $v_{\rm tan}^2 = v^2 - v_{\rm rad}^2$.  Note that this assumes that the centre of mass of the star/disc system is stationary, otherwise the centre of mass velocity would need to be subtracted first.  
If an SPH particle fulfils this criterion it is counted as a `disc' particle and all the terms in equation \ref{eq:quadrature} are used except for the $\left[ v_{ki} \right]_{\rm pres}$ term.  If a particle is not a `disc' particle, it is an `envelope' particle and only the first two terms in equation \ref{eq:quadrature} are used.  In the calculations performed for this paper, the above criterion is found empirically to provide good disc/envelope differentiation even when the disc is very young and/or self-gravitating, containing spiral density waves or particles with eccentric orbits.  Some particles near the outer boundary of the cloud are identified as `disc' particles using this method (because of the use of reflective boundary conditions at the edge of a constant volume sphere and, therefore, their small absolute speeds), but since the dust associated with these particles does not grow significantly during the calculations and the relative dust velocities of such small grains are dominated by Brownian motion (i.e., all other contributions are negligible) this has no adverse affect on the calculations.

Finally, there is an issue in calculations of non-rotating clouds.  Although such calculations do not form a rotating first hydrostatic core or pre-stellar disc, they do form a warm, high-density, pressure-supported first hydrostatic core.  The normalisation of the turbulent velocities, $V_{\rm g} = (3/2)^{1/2}c_{\rm s}$ for the envelope, but only $V_{\rm g} = \alpha_{\rm SS}^{1/2}c_{\rm s}$ in the disc.  It is not reasonable to retain the envelope formulation of the turbulence in a pressure-supported, static first hydrostatic core while assuming much less turbulence in a pressure-supported disc.  Therefore, in addition to the radial/tangential velocity criteria, we also count an SPH particle as a `disc' particle if its radial velocity is less than 10 percent of the local sound speed.  With this additional criterion, the first hydrostatic cores formed in non-rotating clouds are taken to have similar turbulent motions to the first cores/pre-stellar discs produced by rotating clouds.

\subsection{The SPH code}
\label{sec:sph}

The calculations presented here were performed 
using a three-dimensional smoothed particle
hydrodynamics \citep{Lucy1977, GinMon1977} code based on the original 
version of \citeauthor{Benz1990} 
(\citeyear{Benz1990}; \citealt{Benzetal1990}), but substantially
modified as described in \citet*{BatBonPri1995},
\cite{PriBat2007}, and 
parallelised using both OpenMP and MPI.

Gravitational forces between particles and a particle's 
nearest neighbours are calculated using a binary tree.  
The smoothing lengths of particles are variable in 
time and space, set iteratively such that the smoothing
length of each particle 
$h = 1.2 (m/\rho)^{1/3}$ where $m$ and $\rho$ are the 
SPH particle's mass and density, respectively
\citep{PriMon2007}.  To reduce numerical shear viscosity, we use the
\cite{MorMon1997} artificial viscosity
with $\alpha_{\rm_v}$ varying between 0.1 and 1 while $\beta_{\rm v}=2 \alpha_{\rm v}$
\citep[see also][]{PriMon2005}.  The SPH equations are 
integrated using a second-order Runge-Kutta-Fehlberg 
integrator \citep{Fehlberg1969} with individual particle 
time steps (see \citealt*{BatBonPri1995} for further details). 
In addition to the usual SPH time step criteria (e.g., the 
Courant condition), we add a time step constraint for the dust
grain evolution such that the change in $\tilde{\rho}_i$ for any 
dust bin $i$ should not exceed 0.3 of the value of $\tilde{\rho}_i$.

The combined radiative transfer and diffuse ISM model
that was developed by \cite{BatKet2015} is used to model the thermal evolution of the gas.  
For the details of this method, the reader
is directed to that paper, or the brief summary provided by \cite{Bate2019}.  We
used exactly the same method as \cite{Bate2019}, studying only a solar metallicity case.

In this paper, we only follow the collapse of the molecular cloud cores up to the onset of
the second collapse phase due to the dissociation of molecular hydrogen that results in
the formation of the stellar core \citep{Larson1969}.  We do not use sink particles 
\citep{BatBonPri1995} to follow the calculations further.  
Evolution beyond this point is beyond the scope of this
paper, but would be interesting to study in the future.

\subsection{Initial conditions and resolution}
\label{sec:initialconditions}

The initial conditions for the molecular cloud cores consist of unstable Bonnor-Ebert spheres. 
We choose 1-M$_\odot$ clouds with an inner to outer density contrast of 20 and a radius of 5500~au.  These are similar to the molecular cloud cores studied by \cite{BatKet2015, BatLor2017}, except less massive and smaller (their clouds contained 5-M$_\odot$ and had radii of 0.1~pc).  The clouds are contained by spherical, reflective boundary conditions (modelled using ghost particles).  We perform one calculation with a stationary initial cloud, and five calculations with clouds that have uniform rotation initially with rotation rates of $5.83 \times 10^{-14}$, $8.24 \times 10^{-14}$, $1.17 \times 10^{-13}$, $1.65 \times 10^{-13}$  and $2.61 \times 10^{-13}$~rad~s$^{-1}$.  These correspond to ratios of rotational to gravitational potential energy with magnitudes of $\beta=0.0025$, 0.005, 0.1, 0.02, and 0.05, respectively.  The \cite{BatKet2015} combined radiative transfer and diffuse interstellar medium method results in the initial clouds being warmer on the outside (gas temperature $\approx 15$~K than at the centre $\approx 6.5$~K), and gives an overall ratio of thermal energy to the magnitude of the gravitational potential energy of $\alpha=0.42$.

The Bonnor-Ebert density distribution is set up using either $1 \times 10^6$ or $3 \times 10^6$ equal-mass SPH gas particles which are placed on a uniform cubic lattice that is deformed radially to achieve the required density profile.  Our gas resolution is an order of magnitude or more larger than the number of particles required to resolve the local Jeans mass throughout the calculation ($\approx 10^5$ SPH particles per solar mass are required to marginally resolve the minimum Jeans mass; \citealt{BatBur1997, Trueloveetal1997, Whitworth1998}; \citealt*{HubGooWhi2006}).  

We assume an initial dust-to-gas ratio of $\varepsilon = 1/100$.  We begin with a typical \cite{MatRumNor1977} (MRN) dust grain size distribution $N(a) \propto a^{-3.5}$ with $a_{\rm min}=5$~nm and $a_{\rm cutoff}=0.25$~$\mu$m, the same as that used by \cite{HirOmu2009}.  For most of our calculations we use 53 grain size bins with logarithmic spacing from $a_{\rm min}$ to $a_{\rm max}=1$~mm (i.e., 10 bins per decade in size), but we perform one $\beta=0$ calculation using 86 grain size bins with logarithmic spacing from $a_{\rm min}$ to $a_{\rm max}=100$~$\mu$m (i.e., 20 bins per decade in size).  We check the dependence on the number of bins because it is known that using too few bins can result artificially rapid growth \citep*{OhtNakNak1990,Wetherill1990}. In Appendix \ref{appendixA} we show that using either 10 bins or 20 bins per decade in grain size gives similar results.

We assume spherical grains with an intrinsic density of 2.26~g~cm$^{-3}$, appropriate for graphite grains.  We assume that the Epstein drag force is valid for all our calculations.  This requires that $s < (9/4) \lambda$, where $\lambda$ is the gas mean free path, and that the relative velocity between the gas and the dust is much less than the mean thermal velocity of the gas, i.e. $|  \mbox{\boldmath{$v$}}_{\rm D} - \mbox{\boldmath{$v$}}_{\rm G} | \ll v_{\rm th}$ \citep{Weidenschilling1977}.  Both conditions are easily satisfied throughout all of the calculations considered in this paper because, as will be seen below, the dust grains do not grow beyond a few tenths of a millimetre during the calculations even in the densest regions of gas and the dust-gas relative velocities remain low.

\section{Results}
\label{sec:results}

\begin{figure}
\centering \vspace{-0.25cm} \vspace{0cm}
    \includegraphics[width=8.8cm]{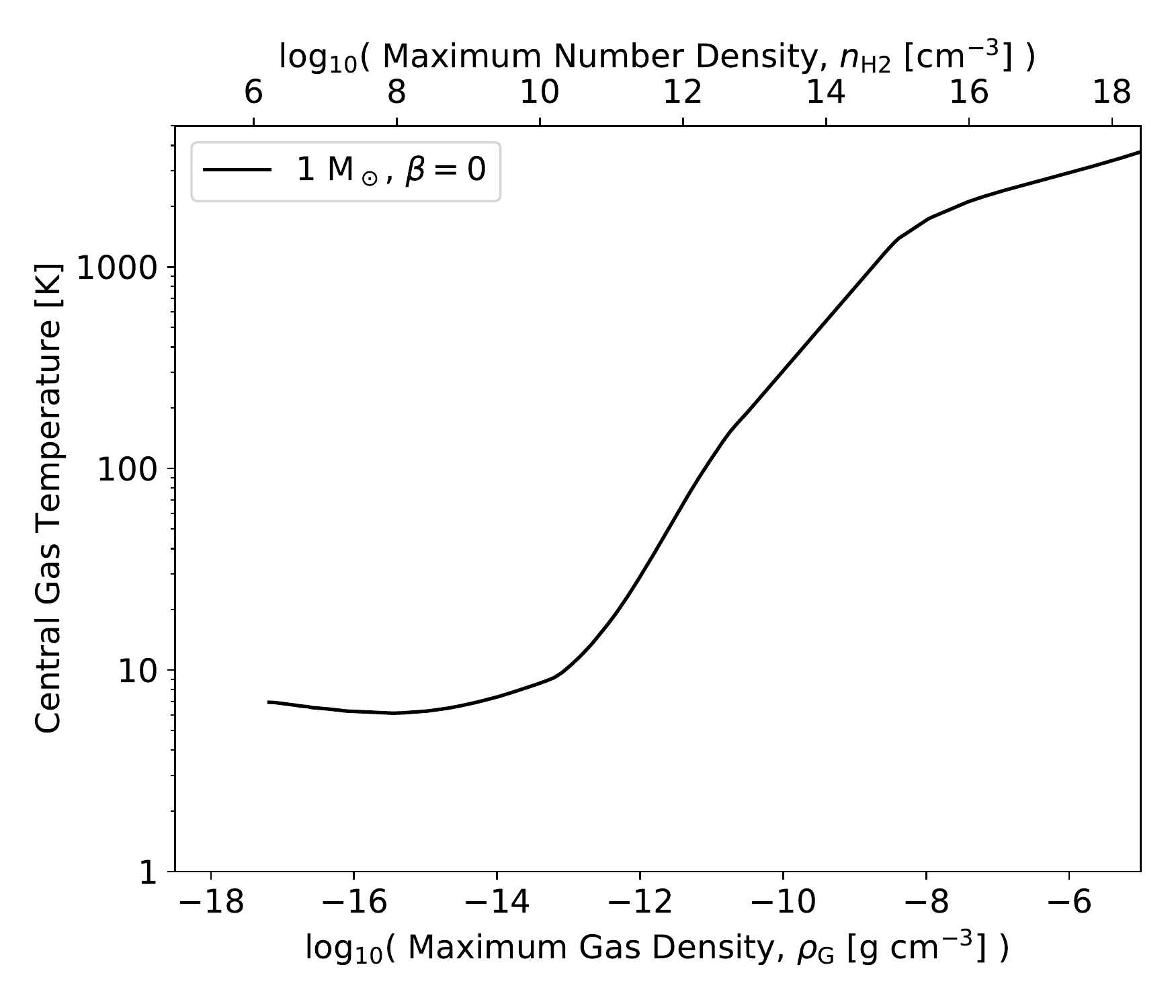}         \vspace{-0.5cm}
\caption{The central gas temperature as a function of the maximum (central) gas density during the collapse of the non-rotating  ($\beta=0$) molecular cloud core. }
\label{fig:dens_temp}
\end{figure}

\subsection{Spherical collapse}
\label{sec:spherical}

We begin by studying dust grain growth during the collapse of the initially static (non-rotating, $\beta=0$) molecular cloud core.  Initially we exclude turbulence as a source of relative grain velocities so that we can compare with the calculations of \cite{HirOmu2009} who studied the coagulation of dust at the centre of collapsing molecular clouds and assumed that the only source of relative grain velocity was due to Brownian motion.  In Fig.~\ref{fig:dens_temp}, we show the evolution of the gas temperature at the centre of the cloud as a function of the maximum density.  The temperature exceeds 1500~K when the number density of molecular hydrogen exceeds $n_\mathrm{H2}\approx 10^{15}$~cm$^{-3}$, so beyond this point any dust grains at the centre of the cloud would be expected to evaporate.

\begin{figure}
\centering \vspace{-0.25cm} \vspace{0cm}
    \includegraphics[width=8.8cm]{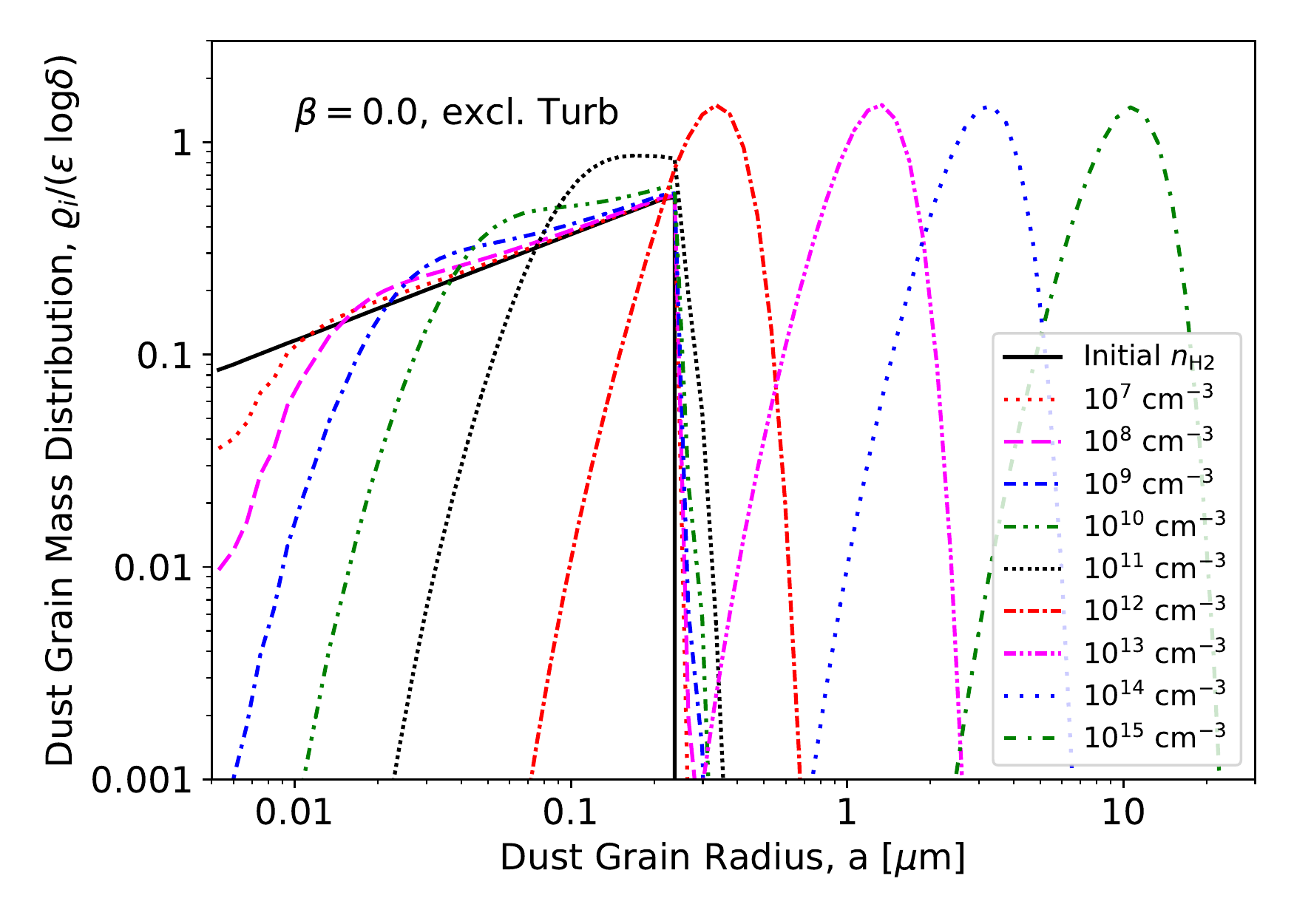} \vspace{-0.5cm}
\caption{The evolution of the dust grain mass distribution function at the centre of the spherically collapsing (non-rotating, $\beta=0$) molecular cloud core as a function of the maximum molecular hydrogen number density, $n_\mathrm{H2}$, during the collapse.  Relative grain velocities are provided by Brownian motion and terminal infall velocities only; the effects of gas turbulence are omitted.  At low densities the smallest grains coagulate reasonably quickly, but the large grains only grow significantly when the hydrogen number density exceeds $n_\mathrm{H2}\approx 10^{11}$~cm$^{-3}$ (gas density $\rho_\mathrm{G} \gtrsim 4 \times 10^{-13}$~g~cm$^{-3}$).}
\label{fig:massdist_beta0}
\end{figure}

In Fig.~\ref{fig:massdist_beta0} we show the evolution of the grain mass distribution at the centre of the collapsing cloud as a function of the molecular hydrogen number density.  The growth of the dust at the centre of the cloud is quite similar to that found by \cite{HirOmu2009}.  At the earliest times and lowest densities, the smallest grains quickly coagulate due to their relatively high relative velocities due to Brownian motion, but the larger grains barely evolve.  The result is that the initial power-law mass distribution ($a^4 N(a) \propto a^{0.5}$) evolves towards a monodisperse population with a peak close to the peak of the initial power-law distribution (see the $n_\mathrm{H2} = 10^{11} - 10^{12}$~cm$^{-3}$ distributions in Fig.~\ref{fig:massdist_beta0}).  As the collapse proceeds further and the grain density increases, the largest grains also begin to coagulate.  This only occurs within the first hydrostatic core that begins to form at number densities of $n_\mathrm{H2} \sim 10^{11}-10^{12}$~cm$^{-3}$ when the radiation begins to be trapped at the centre of the cloud and the temperature increases (Fig.~\ref{fig:dens_temp}) producing a short-lived pressure-supported object \citep{Larson1969}. The result is a peaked distribution of grains that continuously moves to larger sizes/masses.  By the time the central density reaches $n_\mathrm{H2}=10^{15}$~cm$^{-3}$ and the central temperature reaches 1500~K the grains at the centre of the first hydrostatic core have sizes of $s \approx 5-20~\mu$m with the peak at $s \approx 10~\mu$m.

\begin{figure}
\vspace{-0.25cm}
    \includegraphics[height=6.2cm]{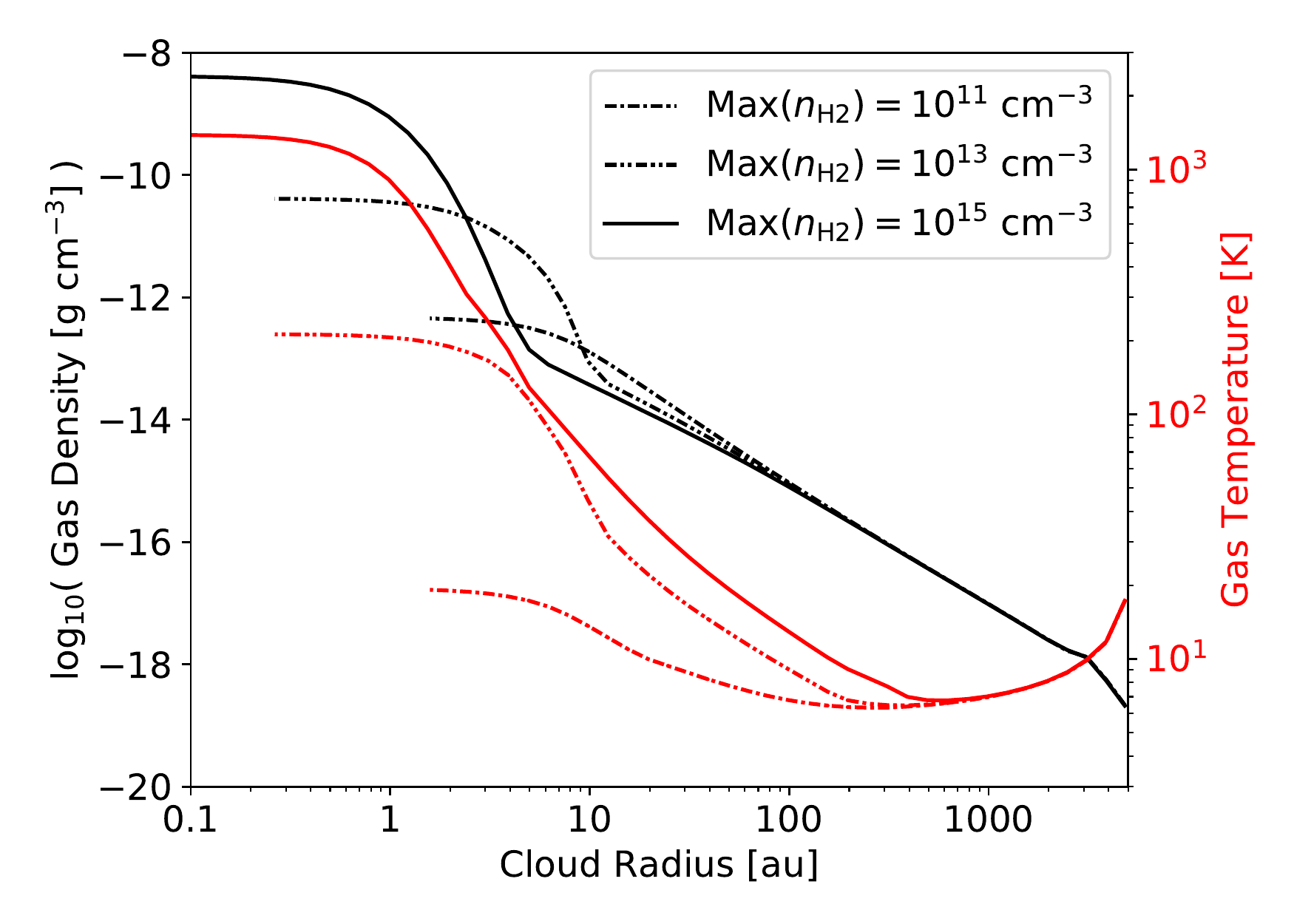}
    \includegraphics[height=6.2cm]{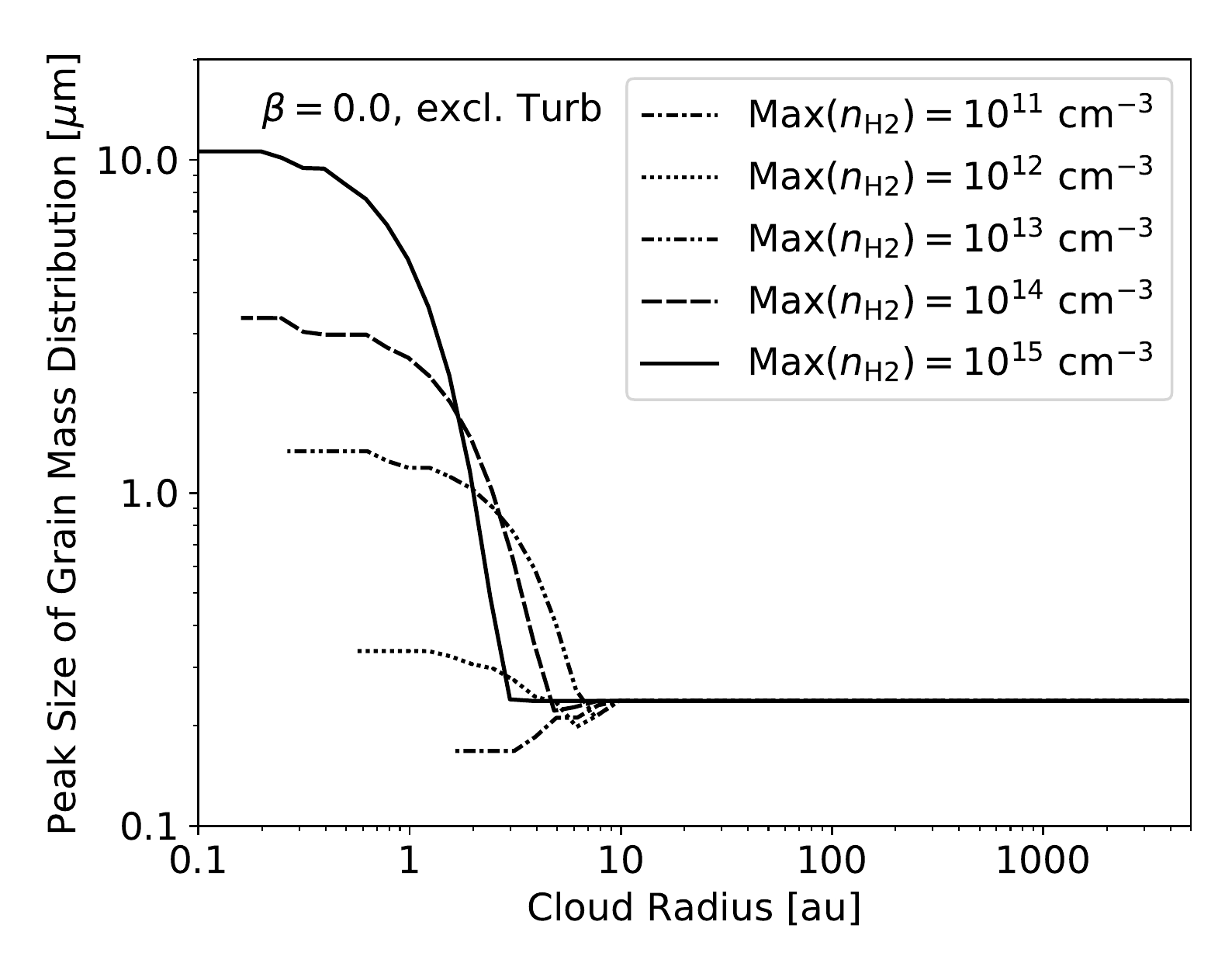} 
\caption{The radial density distributions (top panel, black lines), radial temperature distributions (top panel, red lines), and the peak size of the dust grain mass distributions as functions of radius (bottom panel) at various times during the collapse of the non-rotating  ($\beta=0$) molecular cloud core during the formation of the first hydrostatic core.  Relative grain velocities are provided by Brownian motion and terminal infall velocities only; the effects of gas turbulence are omitted. Both panels provide lines when the maximum molecular hydrogen number density reaches $n_\mathrm{H2} = 10^{11}$~cm$^{-3}$ (dot-dashed lines), $n_\mathrm{H2} = 10^{13}$~cm$^{-3}$ (dot-dot-dashed lines), and $n_\mathrm{H2} = 10^{15}$~cm$^{-3}$ (solid lines), the latter of which is also when the central temperature reaches $\approx$1500~K.  In the lower graph we also plot the peak size for the intermediate times when $n_\mathrm{H2} = 10^{12}$~cm$^{-3}$ (dotted lines) and $n_\mathrm{H2} = 10^{14}$~cm$^{-3}$ (dashed lines). }
\label{fig:peak_size_vs_radius}
\end{figure}

\begin{figure}
\centering \vspace{-0.25cm} \vspace{0cm}
    \includegraphics[height=6.2cm]{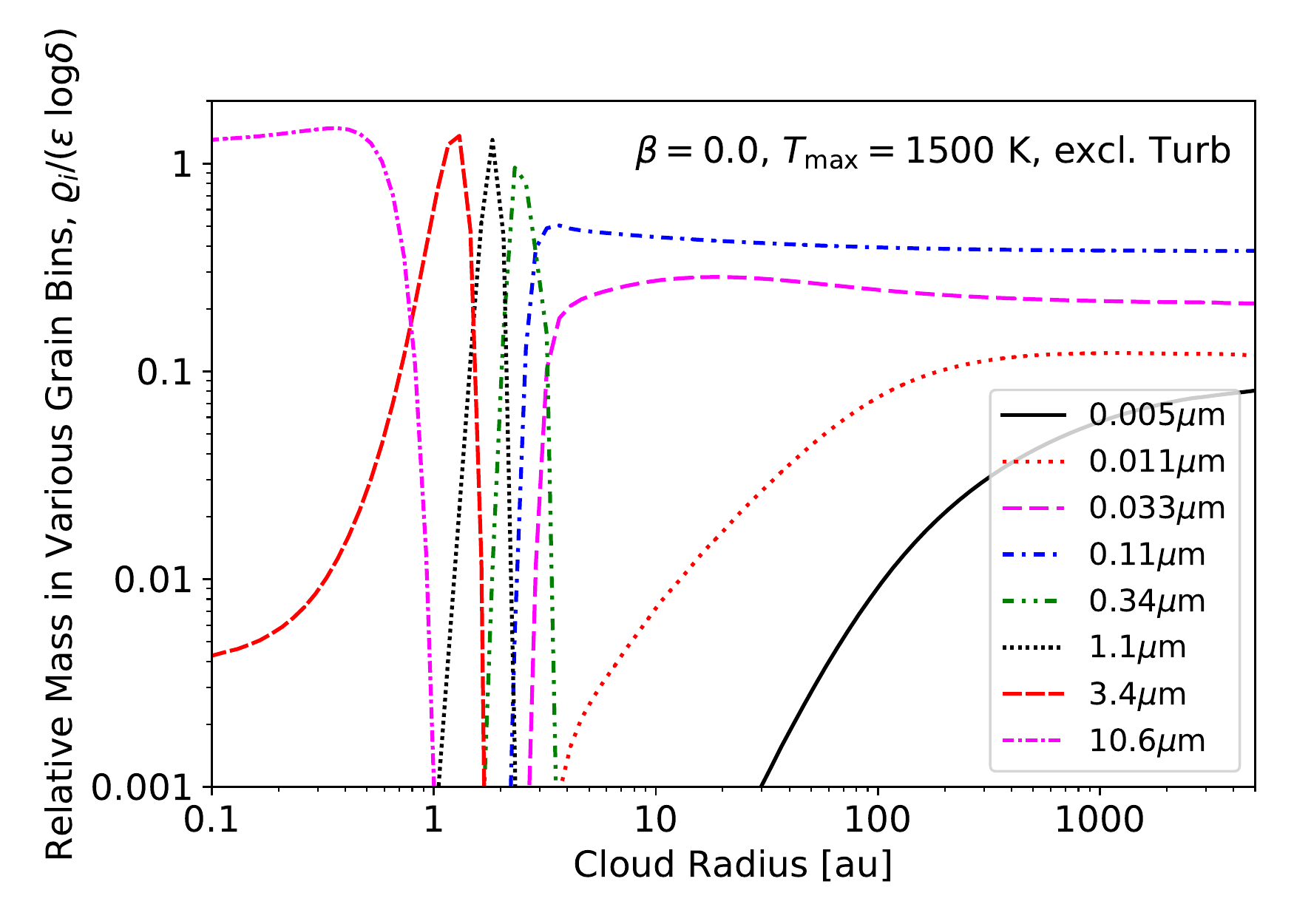}         \vspace{-0.5cm}
\caption{The relative dust masses in selected grain size bins as functions of radius when the non-rotating ($\beta=0$) molecular cloud core has collapsed to produce a first hydrostatic core with a central temperature of 1500~K (maximum molecular hydrogen number density of $n_\mathrm{H2} \approx  10^{15}$~cm$^{-3}$).  Relative grain velocities are provided by Brownian motion and terminal infall velocities only; the effects of gas turbulence are omitted. }
\label{fig:mass_bins}
\end{figure}

Quantitatively, the results differ slightly to those presented by \cite{HirOmu2009}.  For example, although their distribution at $n_\mathrm{H2} = 10^{10}$~cm$^{-3}$ for their solar-metallicity case is very similar to our distribution at the same maximum density, a short time later during the collapse their peak size is $s \approx 0.45~\mu$m at $n_\mathrm{H2} = 10^{13}$~cm$^{-3}$ in their solar-metallicity case, compared to our peak value of $s \approx 1~\mu$m at the same maximum density.  Small quantitative differences are to be expected for two main reasons.  First, they have an initial dust-to-gas ratio of 0.006.  Since this is lower, it is expected that the grain growth should be slightly slower.  Second, they did not perform a full hydrodynamical calculation to obtain their evolution of density versus time; they used an analytic model based on the central free-fall time.  Therefore, the time evolution of the density and temperature that they used will be somewhat different to our model, which will lead to a somewhat different dust growth rate at the centre of the cloud for the two models.

In Fig.~\ref{fig:peak_size_vs_radius}, we show that the large grains are confined to the first hydrostatic core that forms at the centre of the collapsing cloud by plotting the gas density and temperature, and the peak size of the grain mass distribution versus radius when the central density reaches $n_\mathrm{H2}=10^{11}$, $10^{13}$, and $10^{15}$~cm$^{-3}$.  Clearly, throughout the entire collapsing envelope ($r \gtrsim 10$ au), the peak grain size is still equal to the initial peak.  Larger grains only exist within the first hydrostatic core where the grain density is high enough to produce significant growth in the relatively short amount of time that is available.

Another way to examine the grain growth that has occurred during the collapse and the first hydrostatic core phase is to plot the relative amounts of the total dust mass that is contained in various dust size intervals (i.e., dust bins).  We plot these relative dust masses for some of the size bins versus radius in Fig.~\ref{fig:mass_bins} when the central temperature reaches 1500~K and the maximum density is $n_\mathrm{H2} \approx 10^{15}$~cm$^{-3}$ (this is just before the second collapse phase would begin, in which the interior of the first hydrostatic core collapses to form the stellar core due to molecular hydrogen dissociation; \citealt{Larson1969}).  At large radius ($r \gtrsim 1000$~au), the grain size distribution is almost identical to the  MRN distribution that was used for the initial cloud: the peak of the grain distribution is still at $s \approx 0.25~\mu$m, most of the mass is contained in the larger grains, but grains with sizes down to 5~nm still exist in large numbers.  However, the smallest grains grow fastest because of their comparatively large relative velocities due to Brownian motion and their comparatively high number density.  The smallest grains diminish with decreasing radius between $r \approx 1000$ and $r\approx 50$ au and almost completely disappear within 30 au.  Grains with sizes $s \approx 10-20$~nm diminish at radii between $r\approx 50$ au and $r\approx 5$ au.  So the smallest grains are coagulated with each other and with larger grains, but the largest initial grains ($s \approx 30-250$~nm) do not begin to grow substantially unless they are within the first hydrostatic core ($r\lesssim 5$~au).  Here, grains grow very quickly due to their high number densities and the higher temperatures (which increase the Brownian velocities).  In the innermost region of the first hydrostatic core ($r \lesssim 1$~au), most of the dust mass is contained in dust grains with sizes of $s \approx 5-20~\mu$m just before the second collapse phase begins.

The relative grain velocities that allow grains to collide and grow in the spherically-collapsing cloud are primarily due to Brownian motion rather than terminal velocities in the infalling envelope.  This is because the initial grains are small, so their Brownian speeds are large and their terminal velocities are very low.  Turning off the terminal velocity contribution to the relative grain velocities (equation \ref{eq:terminalv}) makes almost no difference to the results discussed above and presented in Figs.~\ref{fig:massdist_beta0}--\ref{fig:mass_bins}.  The first two figures are indistinguishable.  In Fig.~\ref{fig:mass_bins}, a small  difference becomes apparent deep within the first core in that without the terminal velocity contribution there is slightly more mass in 10~$\mu$m grains within 0.3~au and slightly less mass in these grains between 0.3 and 1~au (less than a factor of two).   Conversely, the relative mass in 3.4~$\mu$m grains is slightly higher within $\approx 0.8$~au (up to a factor of $\approx 3$) without the terminal velocity contributions.  The other distributions are indistinguishable.  As grains become larger ($s \gtrsim 3~\mu$m), the effect of Brownian motions diminishes but they are less coupled to the gas so the terminal velocities increase.  Nevertheless, the growth of the grains is completely dominated by Brownian motions throughout non-rotating calculation in the absence of stirring due to gas turbulence.

\begin{figure}
\centering \vspace{-0.25cm} \vspace{-0.1cm}
    \includegraphics[height=6.2cm]{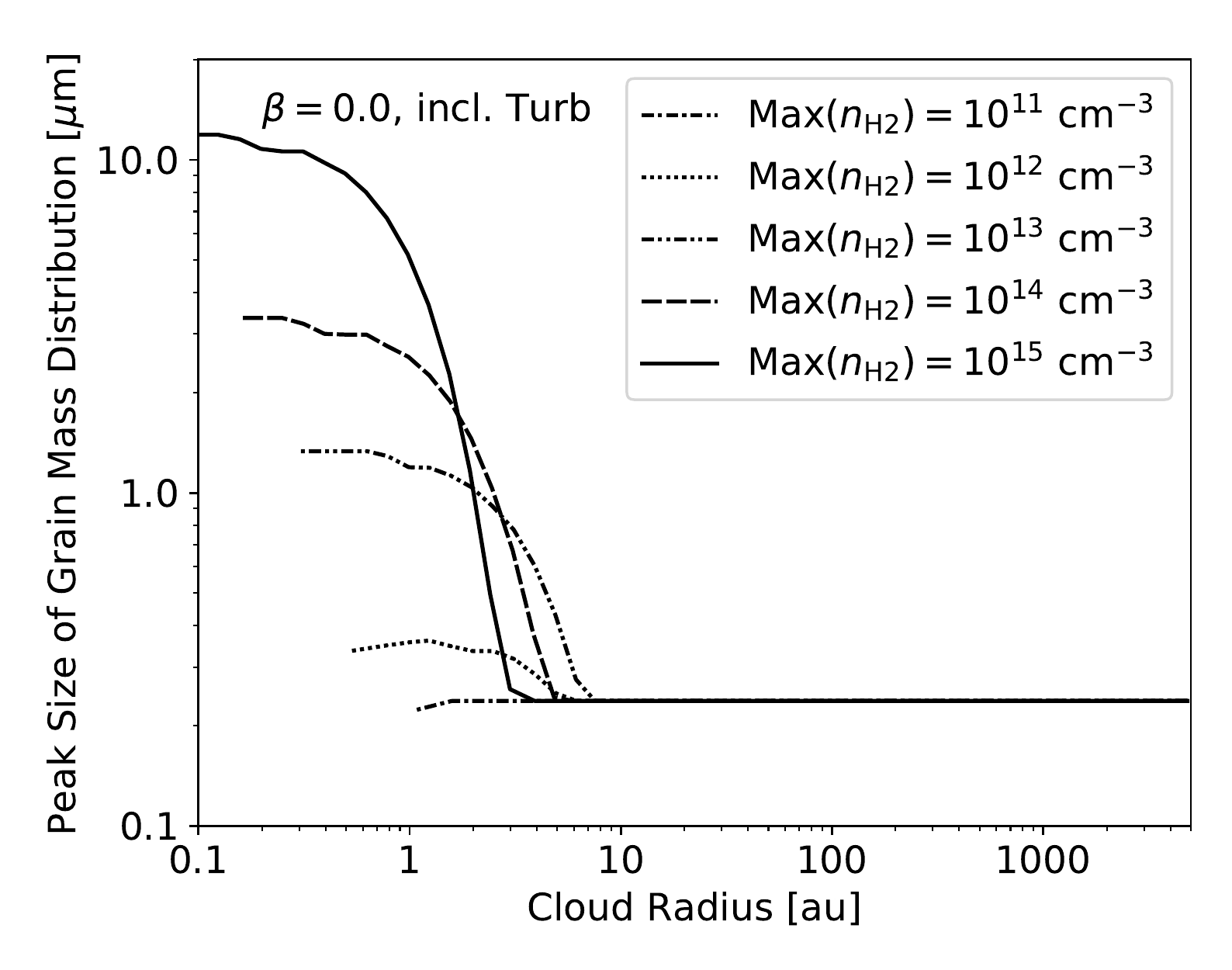}        
    \includegraphics[height=6.2cm]{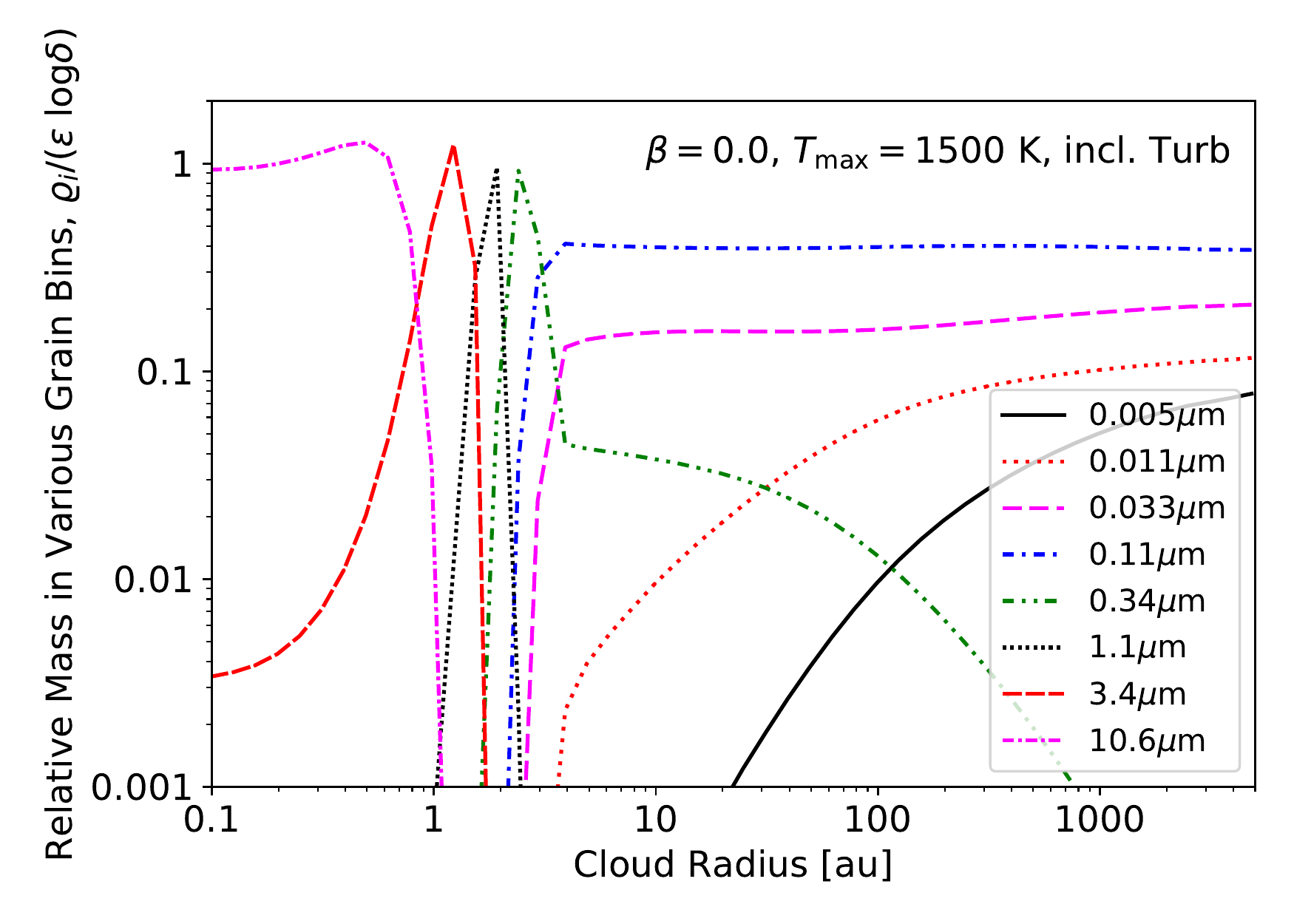}         \vspace{-0.5cm}
\caption{The equivalents of Figs.~\ref{fig:peak_size_vs_radius} and \ref{fig:mass_bins}, but for the collapse of a non-rotating  ($\beta=0$) molecular cloud core that includes relative grain velocities driven by gas turbulence, as well as Brownian motion and terminal infall velocities.  The upper panel shows the peak size of the dust grain mass distributions as functions of radius at various times during the collapse and the formation of the first hydrostatic core. The lower panel provides the relative dust masses in selected grain size bins as functions of radius when the central temperature reaches 1500~K.  The main effect of the additional turbulence-driven grain velocities is to produce a significant population of dust grains with sizes slightly larger ($s\approx 0.3-0.5$~$\mu$m) than the cut-off in the initial MRN grain distribution ($s= 0.25$~$\mu$m). }
\label{fig:mass_bins_turb}
\end{figure}

How does including turbulence-driven relative grain velocities affect the dust evolution?  In Fig.~\ref{fig:mass_bins_turb} we show the equivalents of Figs.~\ref{fig:peak_size_vs_radius}--\ref{fig:mass_bins} but including the effect of turbulence (Section \ref{sec:turbulence}).  Comparing the equivalent figures, the depletion of the very smallest grains ($s \lsim 0.01$~$\mu$m) in the envelope is almost identical the case without turbulence.  For these very small dust grains, the Brownian motions and turbulence-induced motions are of similar magnitudes.   Slightly larger dust grains ($s \approx 0.03$~$\mu$m) suffer a little more depletion when the turbulence-induced motions are included while the distribution of $s\approx 0.1$~$\mu$m grains is similar.  The main difference when turbulent stirring is included is that grains slightly larger than the initial cut-off size ($s = 0.25$~$\mu$m) are produced with a significant abundance within the collapsing envelope, exterior to the first hydrostatic core (compare the $s \approx 0.34$~$\mu$m distributions).  However, despite this, the production of grains significantly larger than those contained in the initial population ($s \gsim 0.5$~$\mu$m) is negligible until the first hydrostatic core is formed, whereupon the much higher grain number densities and temperatures produce much more rapid grain growth.  The population of large grains within the first hydrostatic core does not depend significantly on whether turbulent-driven relative grain velocities are included or not, as long as the turbulence within the first core is assumed to be weak (i.e., similar to that expected in a disc).

\begin{figure}
\centering \vspace{-0.25cm} \vspace{-0.1cm}
    \includegraphics[height=6.8cm]{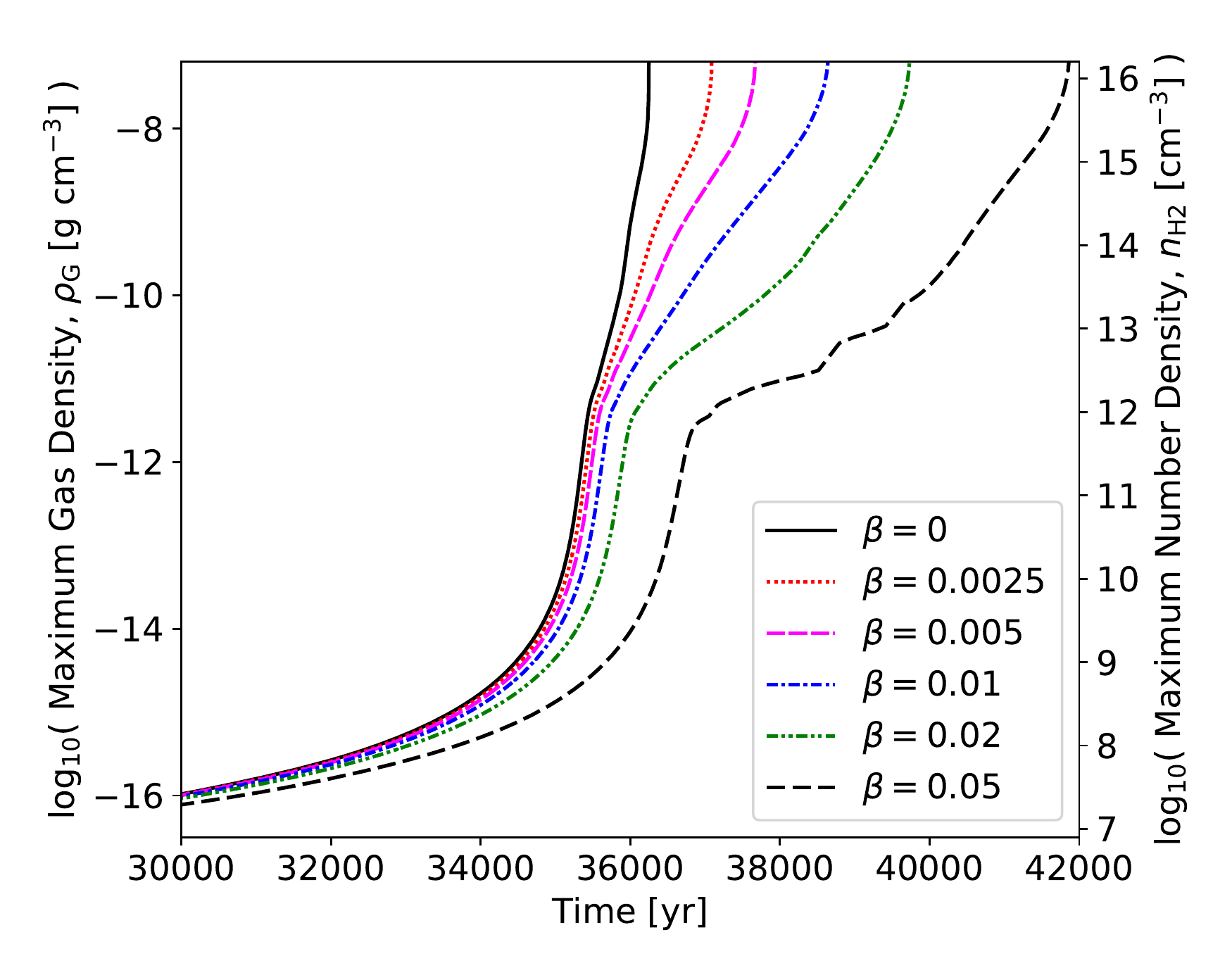}         \vspace{-0.5cm}
\caption{The maximum (central) density as a function of time for six molecular cloud cores with different rotation rates.  The first hydrostatic core phase lasts longer with more rotational support (maximum density $\rho \approx 10^{-12} - 10^{-8}$~g~cm$^{-3}$ or $n_{\rm H2} \approx 10^{12} - 10^{16}$~cm$^{-3}$). }
\label{fig:density_time}
\end{figure}

\begin{figure*}
\vspace{-0.5cm}
    \includegraphics[width=7.5cm]{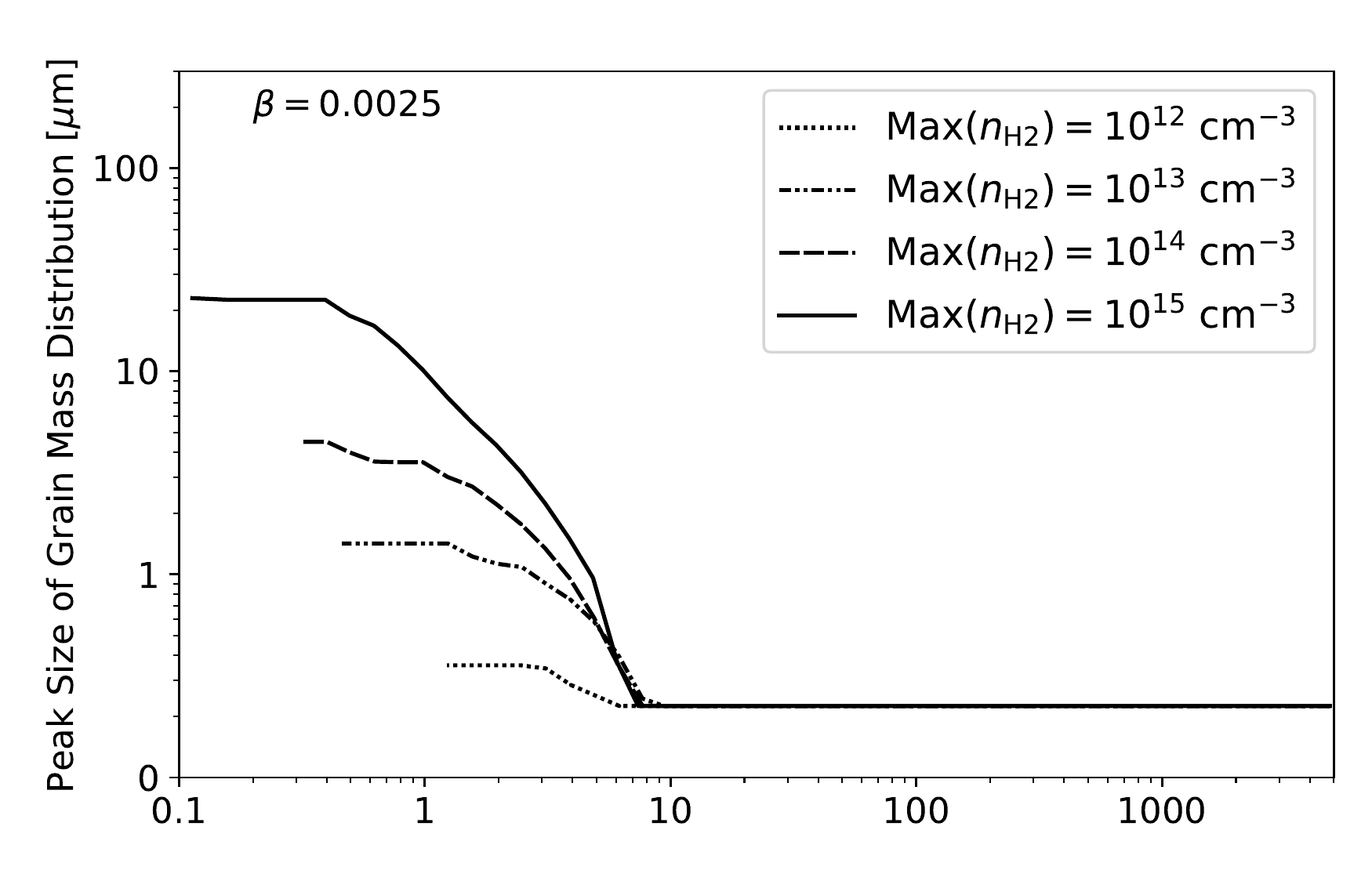}\hspace{0.5cm} 
    \includegraphics[width=7.5cm]{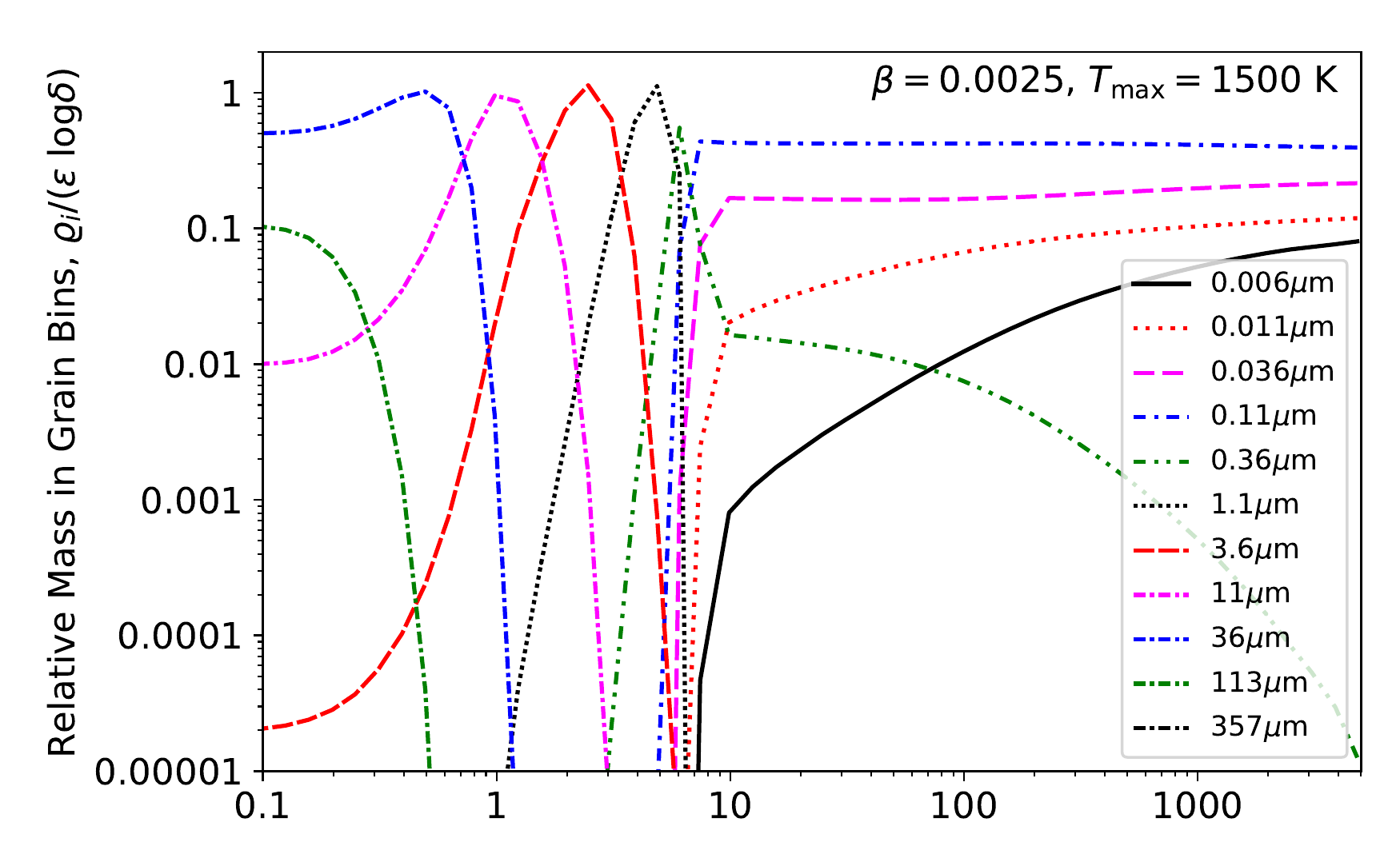}         \vspace{-0.3cm}

    \includegraphics[width=7.5cm]{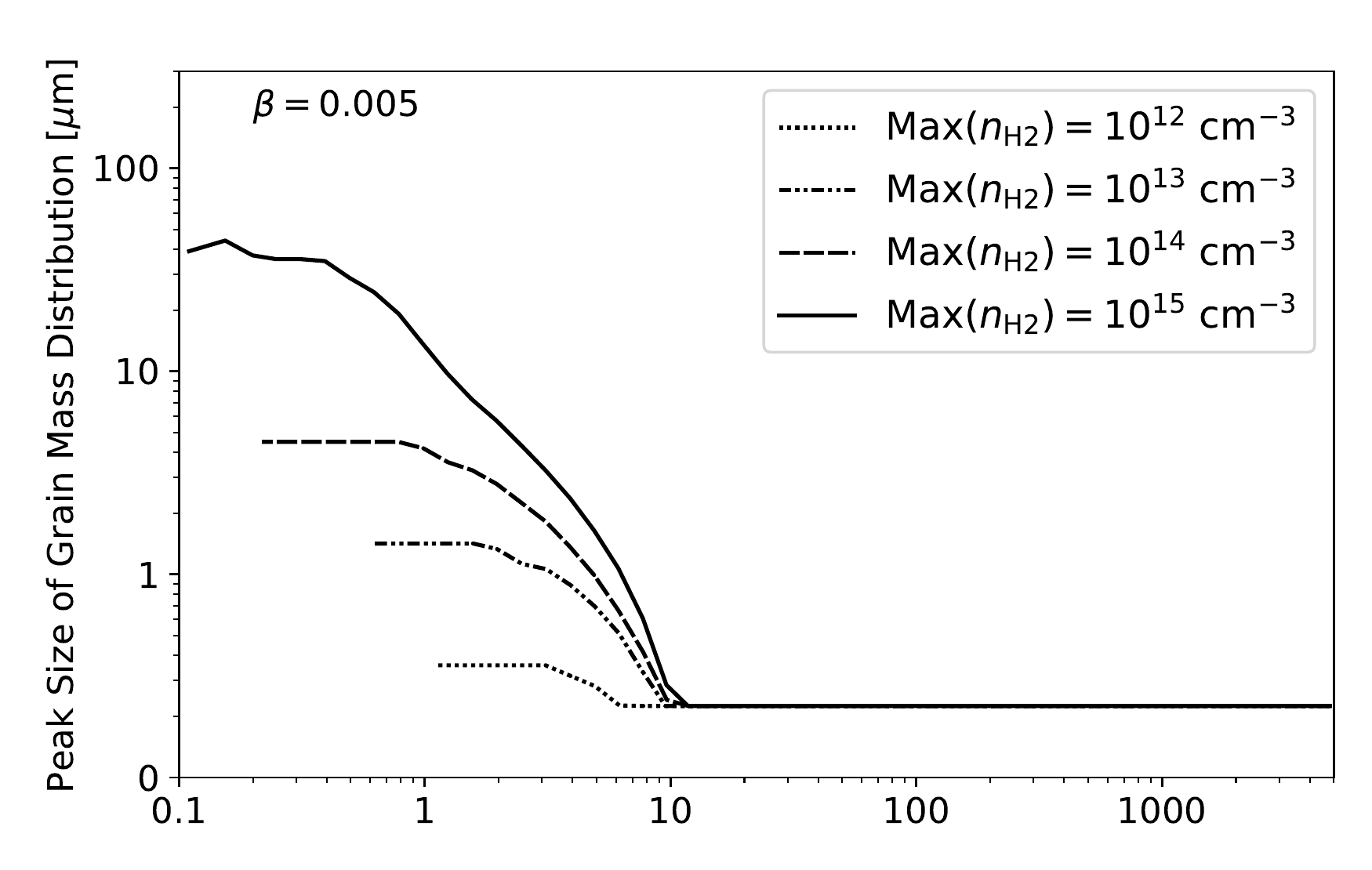}\hspace{0.5cm} 
    \includegraphics[width=7.5cm]{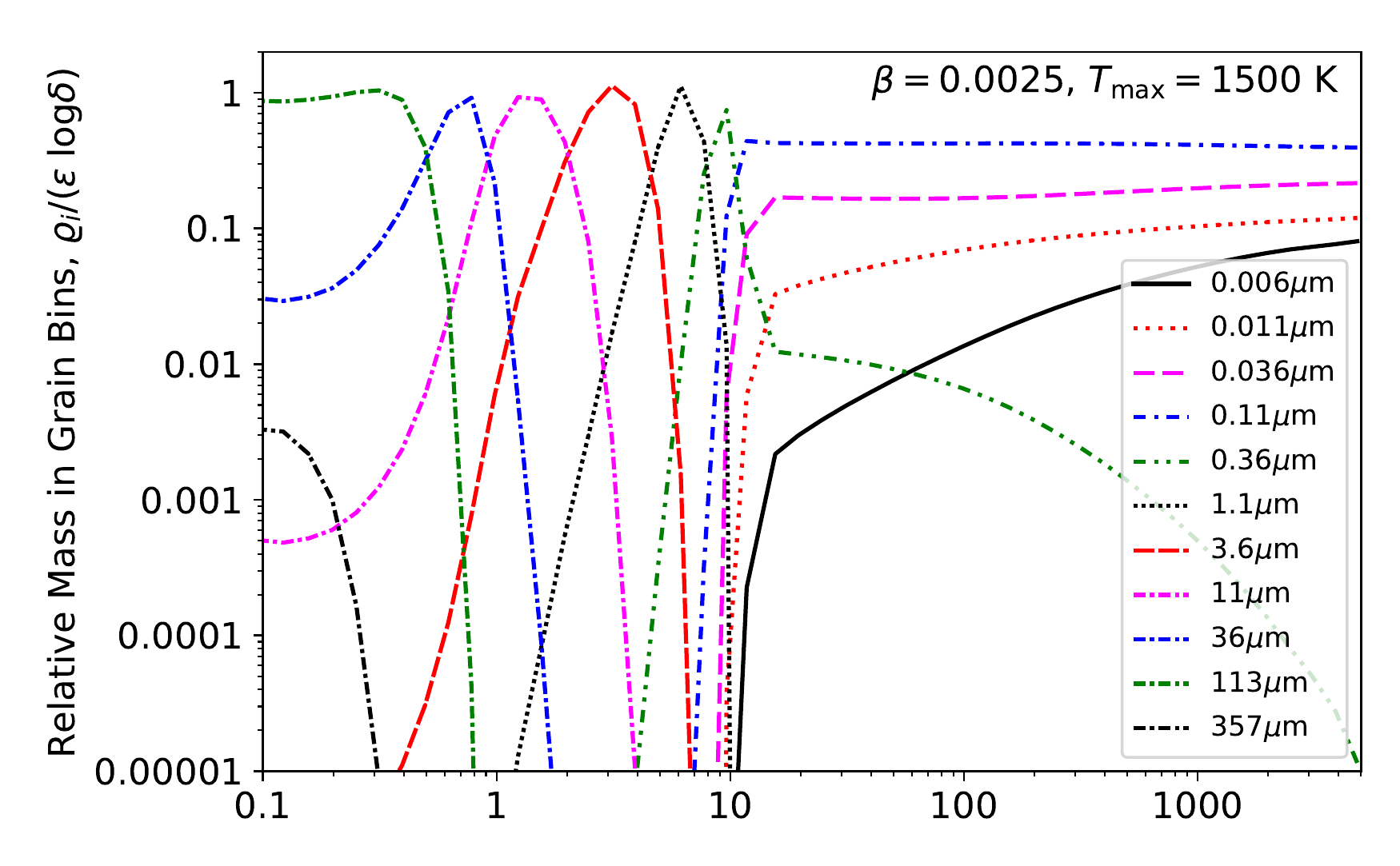}         \vspace{-0.3cm}

    \includegraphics[width=7.5cm]{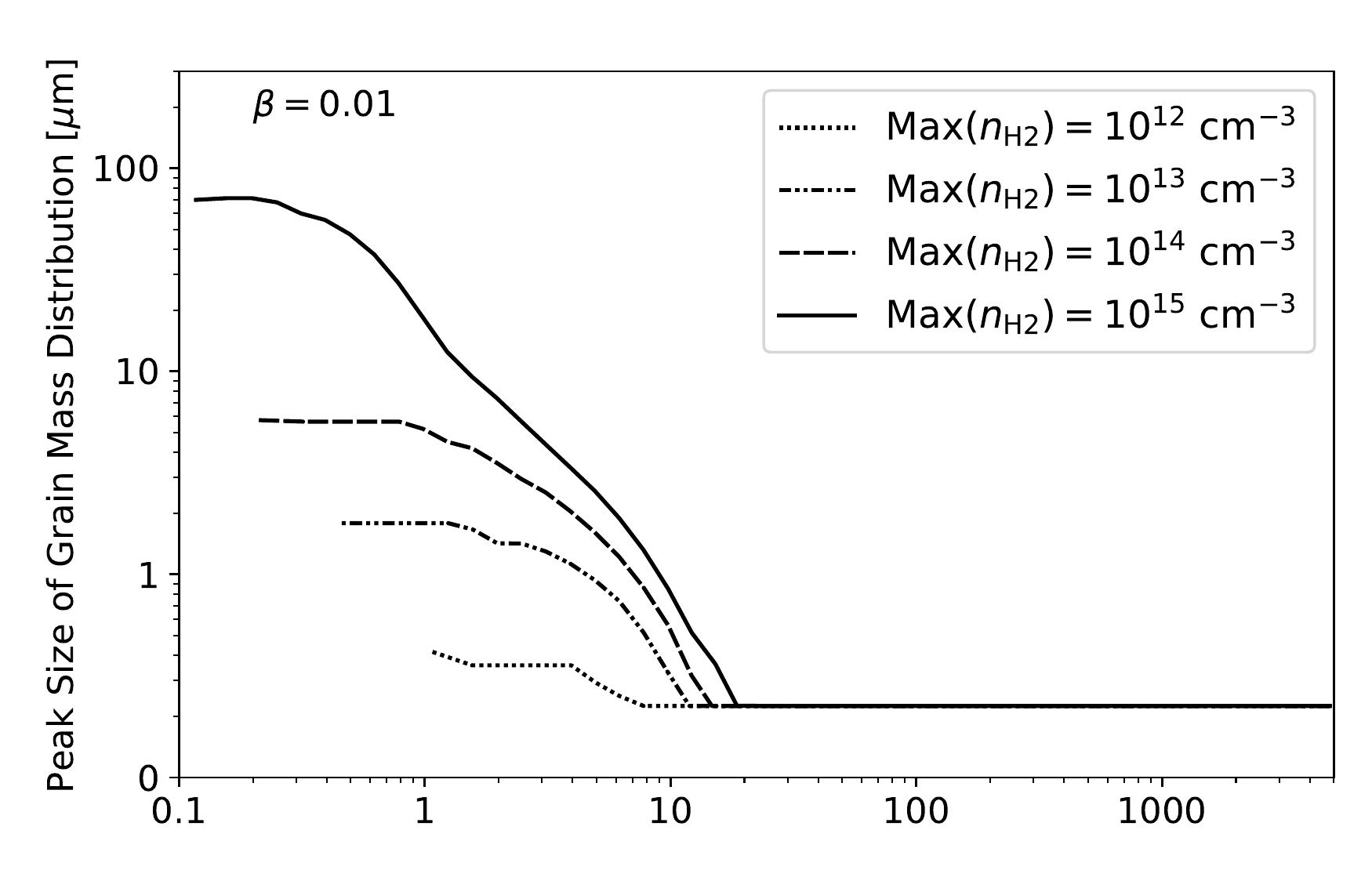}\hspace{0.5cm} 
    \includegraphics[width=7.5cm]{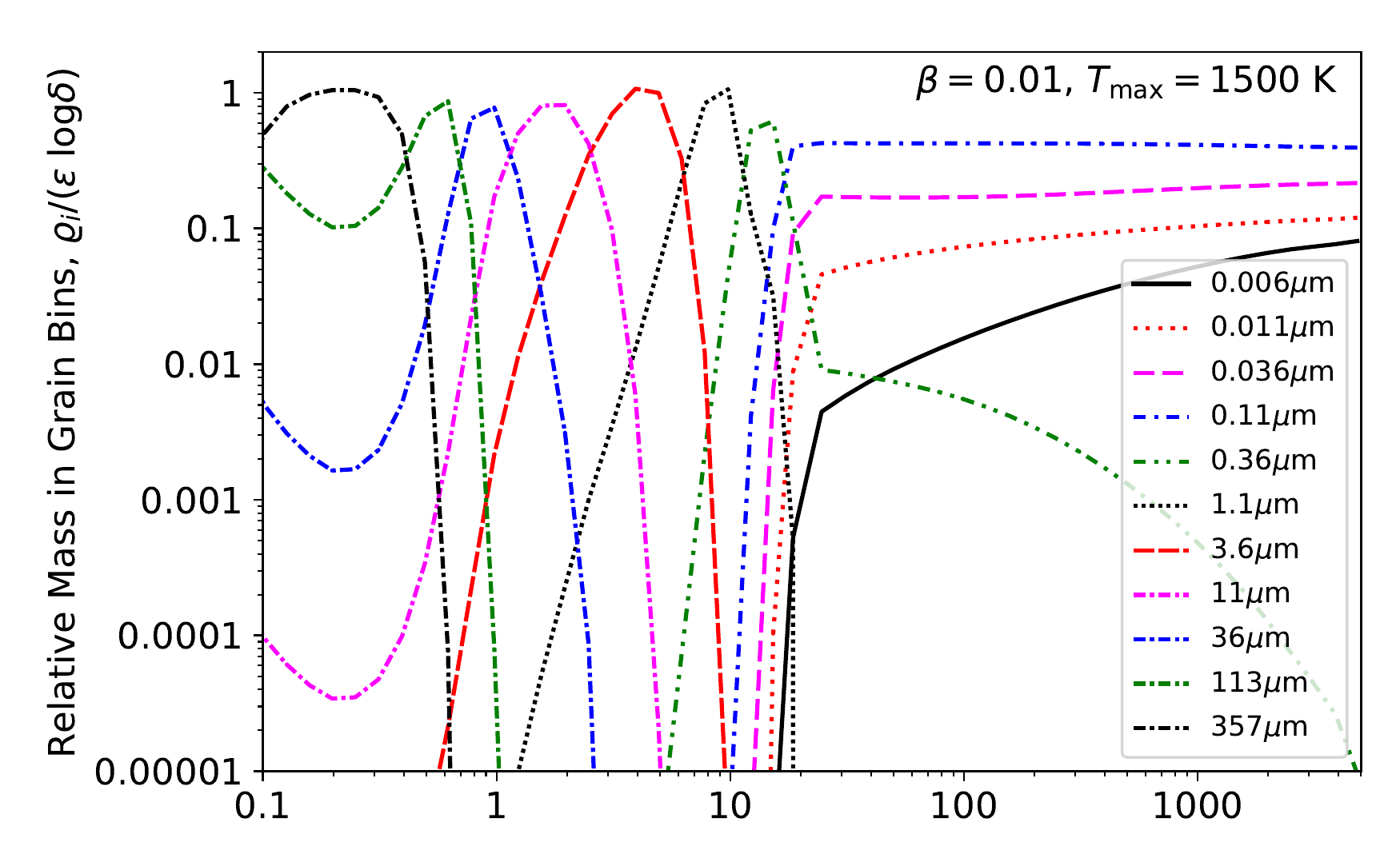}         \vspace{-0.3cm}

    \includegraphics[width=7.5cm]{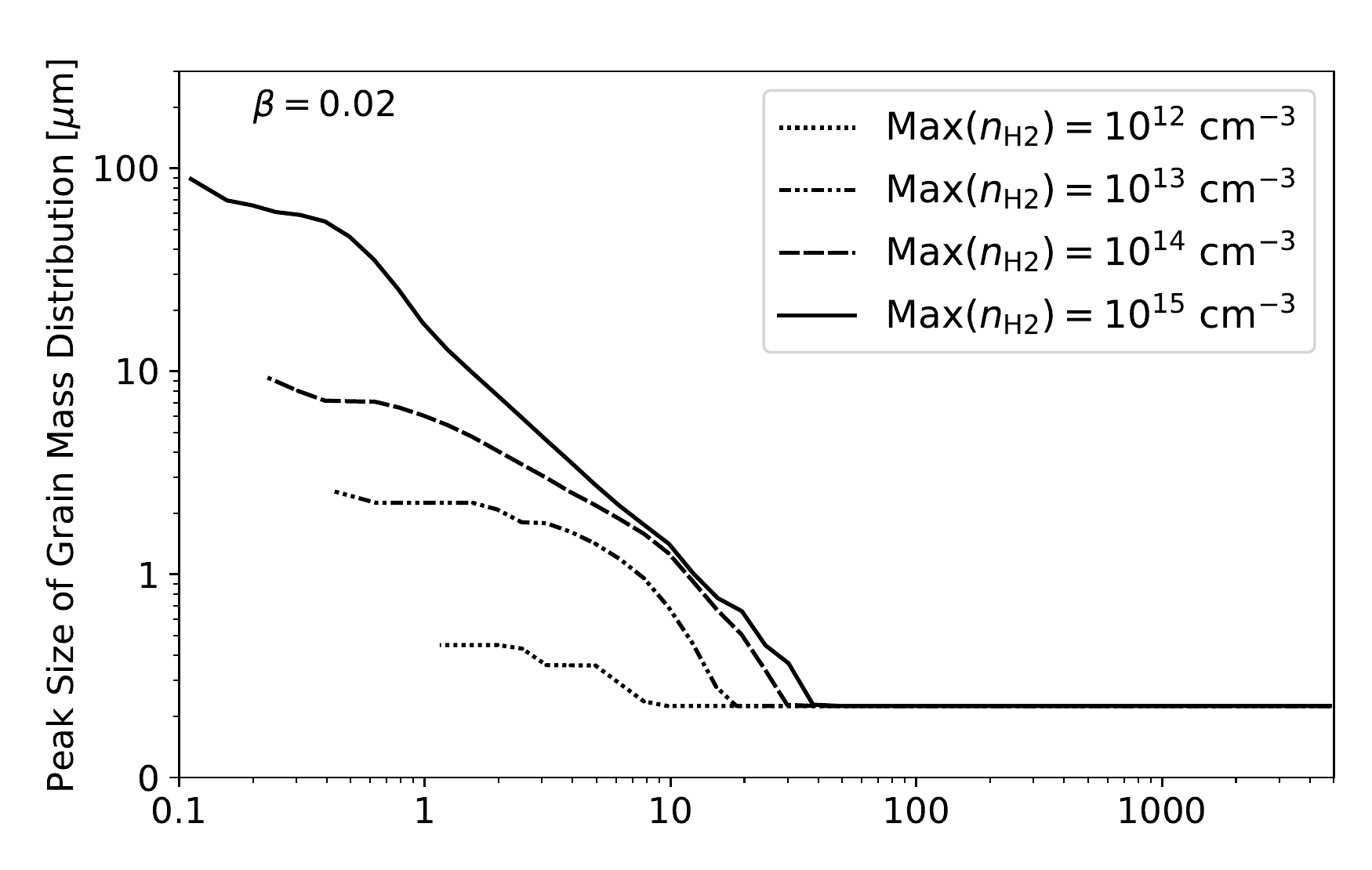}\hspace{0.5cm} 
    \includegraphics[width=7.5cm]{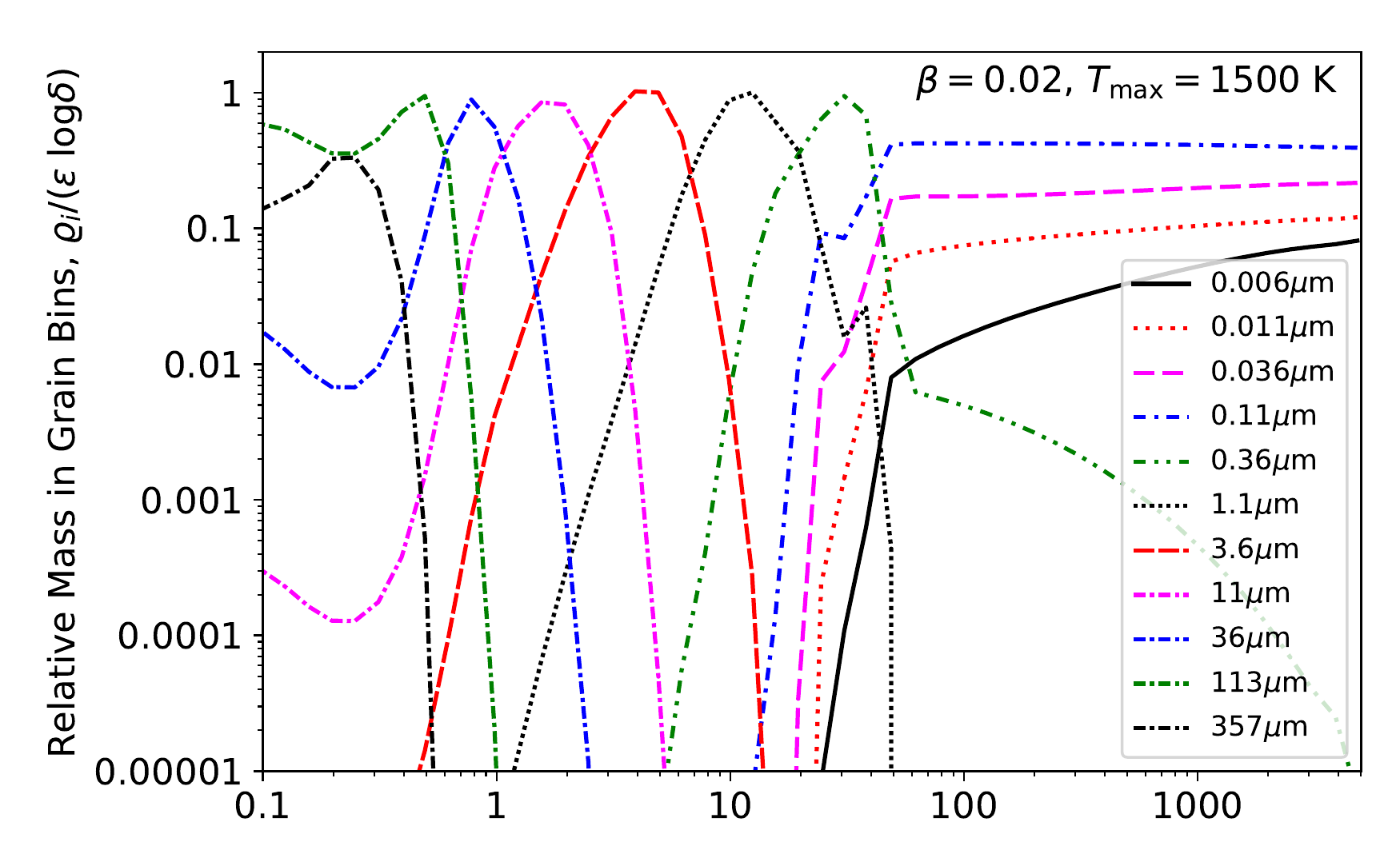}         \vspace{-0.3cm}
    
    \includegraphics[width=7.5cm]{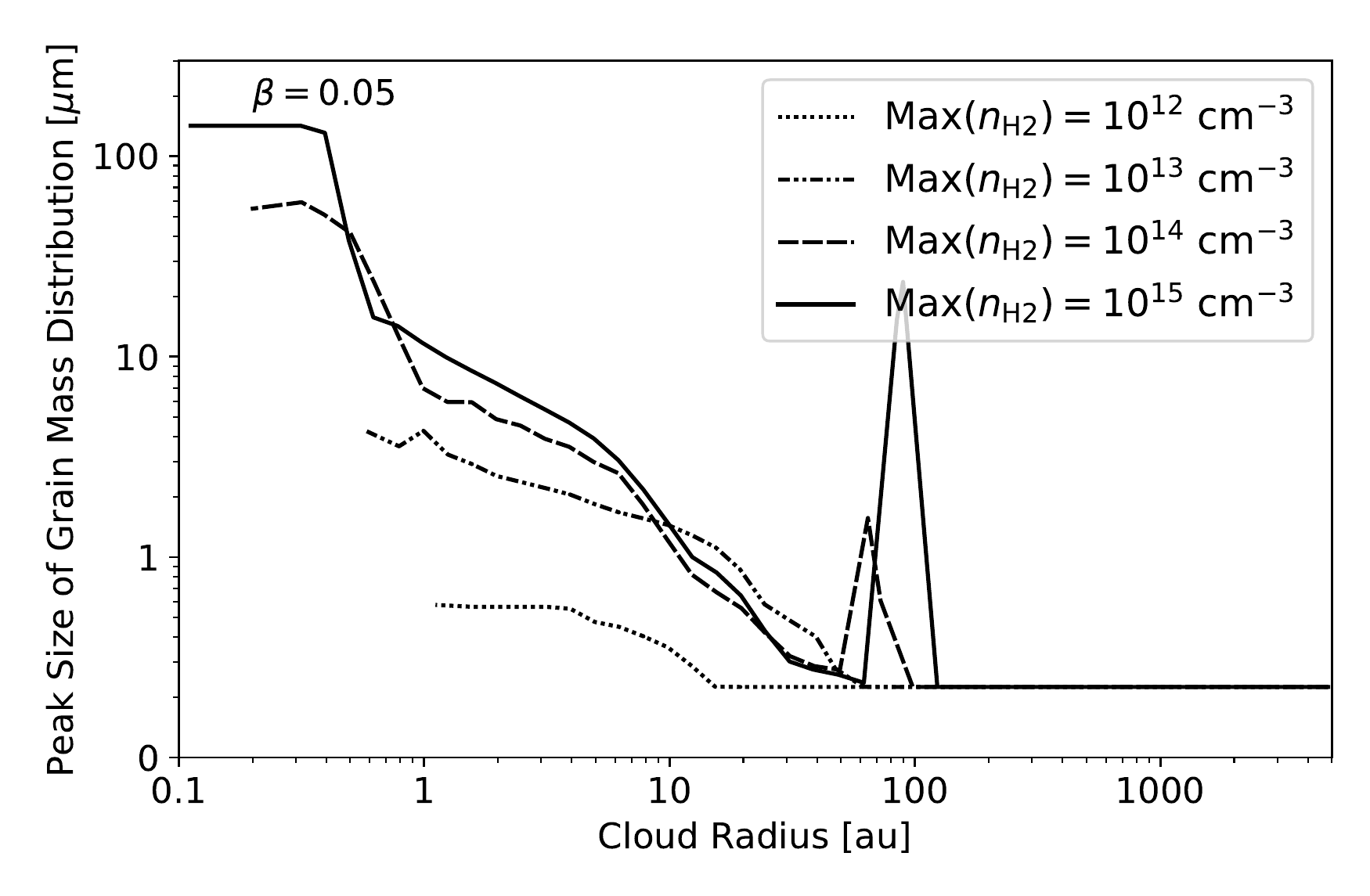}\hspace{0.5cm} 
    \includegraphics[width=7.5cm]{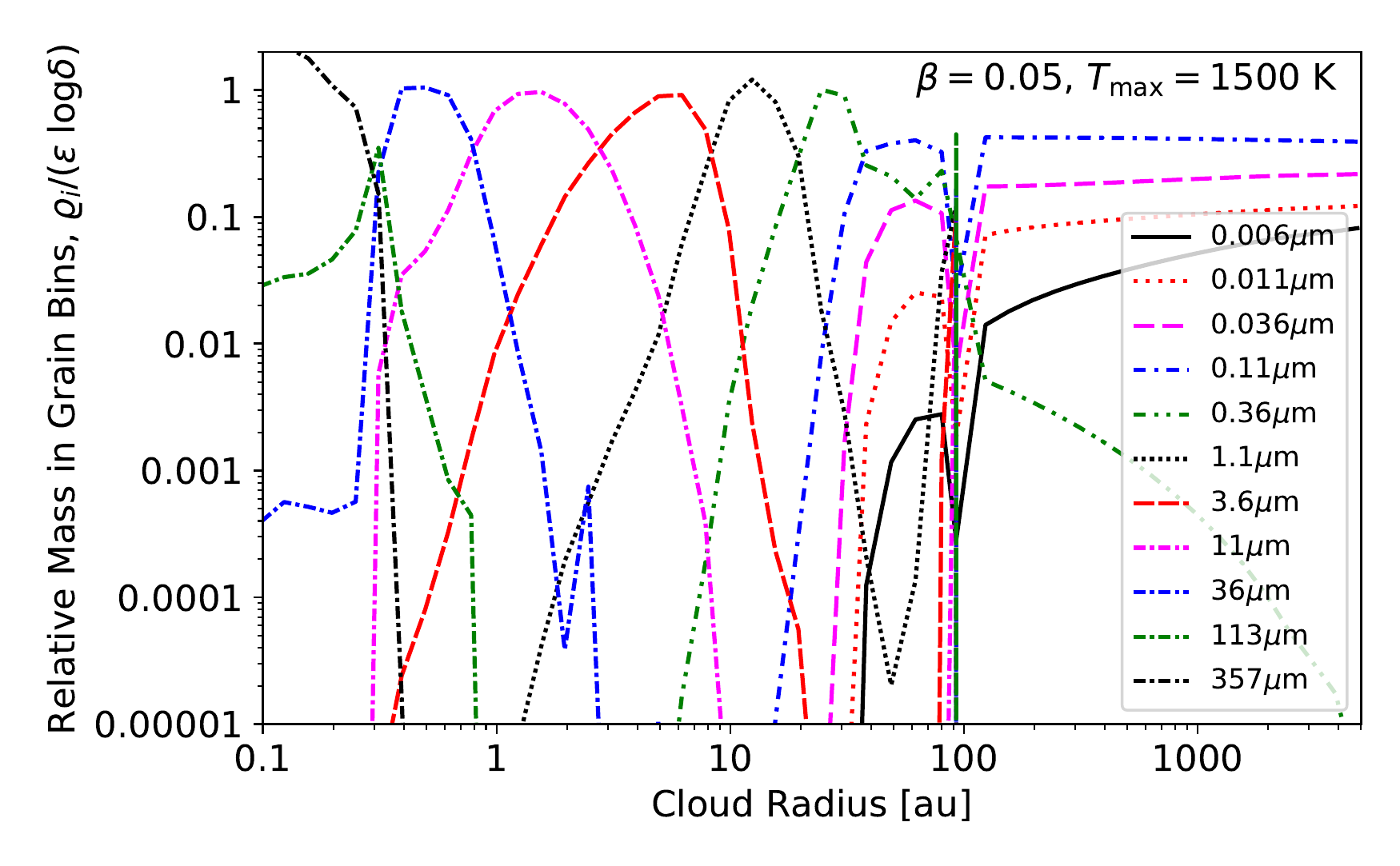}         \vspace{-0.3cm}
    
\caption{Left panels: the peak sizes of the dust grain mass distributions as functions of radius near the mid-plane at four times (denoted by the maximum number density, $n_{\rm H2}$) during the evolution of the rotating molecular cloud cores ($\beta=0.0025, 0.005, 0.01, 0.02, 0.05$ from top to bottom) after the formation of the first hydrostatic cores.   Right panels: the relative dust masses in selected grain size bins as functions of radius near the mid-plane when the maximum temperature in the same five models has reached 1500~K (which is also when the maximum molecular hydrogen number density is $n_\mathrm{H2} \approx  10^{15}$~cm$^{-3}$).  The features at $r\approx 100$~au in the $\beta=0.05$ case are due to the formation of a fragment in the disc (see Section \ref{sec:projected}).}
\label{fig:rot_mass_bins}
\end{figure*}

\begin{figure*}
\vspace{-0.5cm}
    \includegraphics[width=7.7cm]{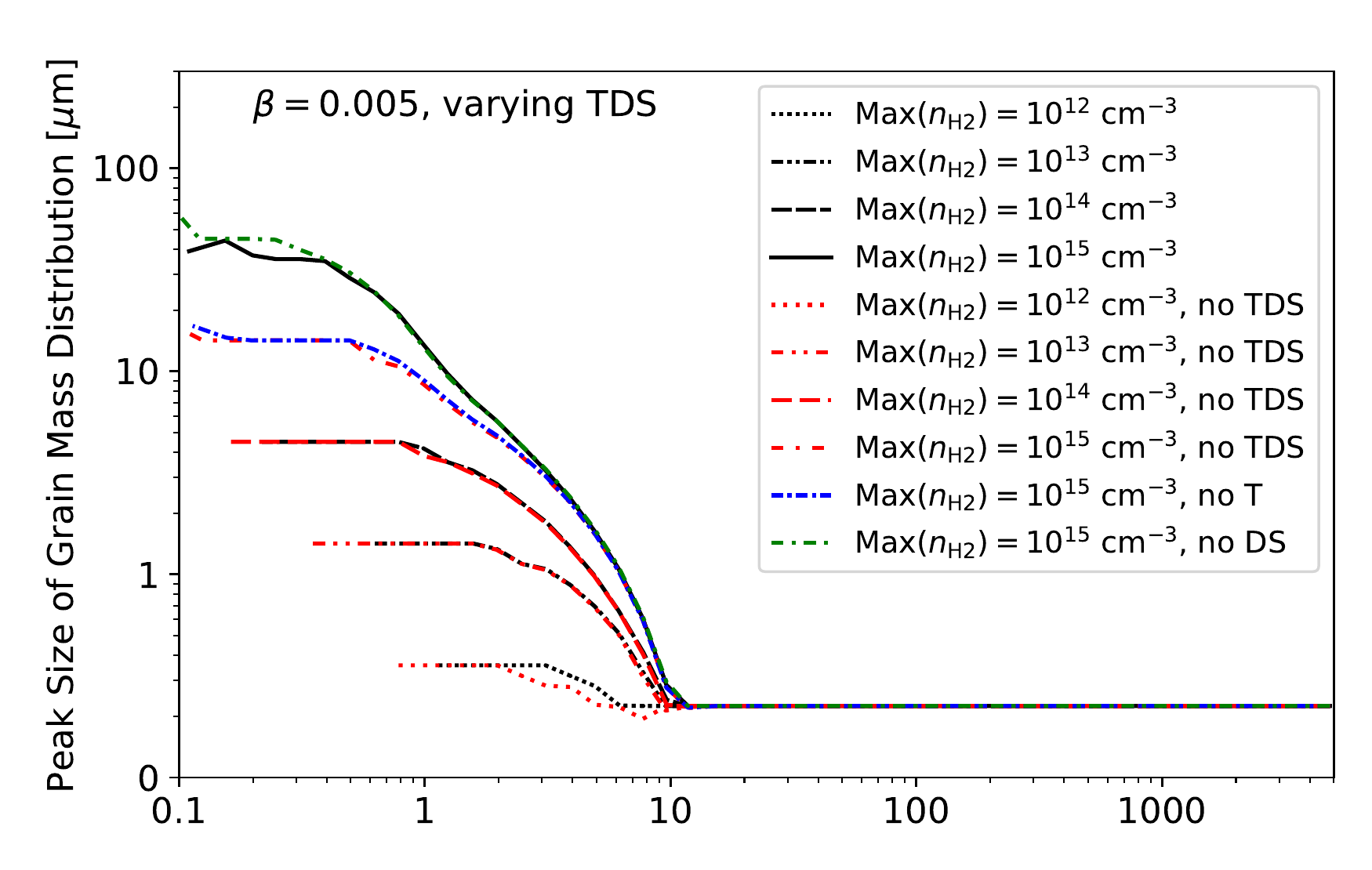}\hspace{0.5cm} 
    \includegraphics[width=7.7cm]{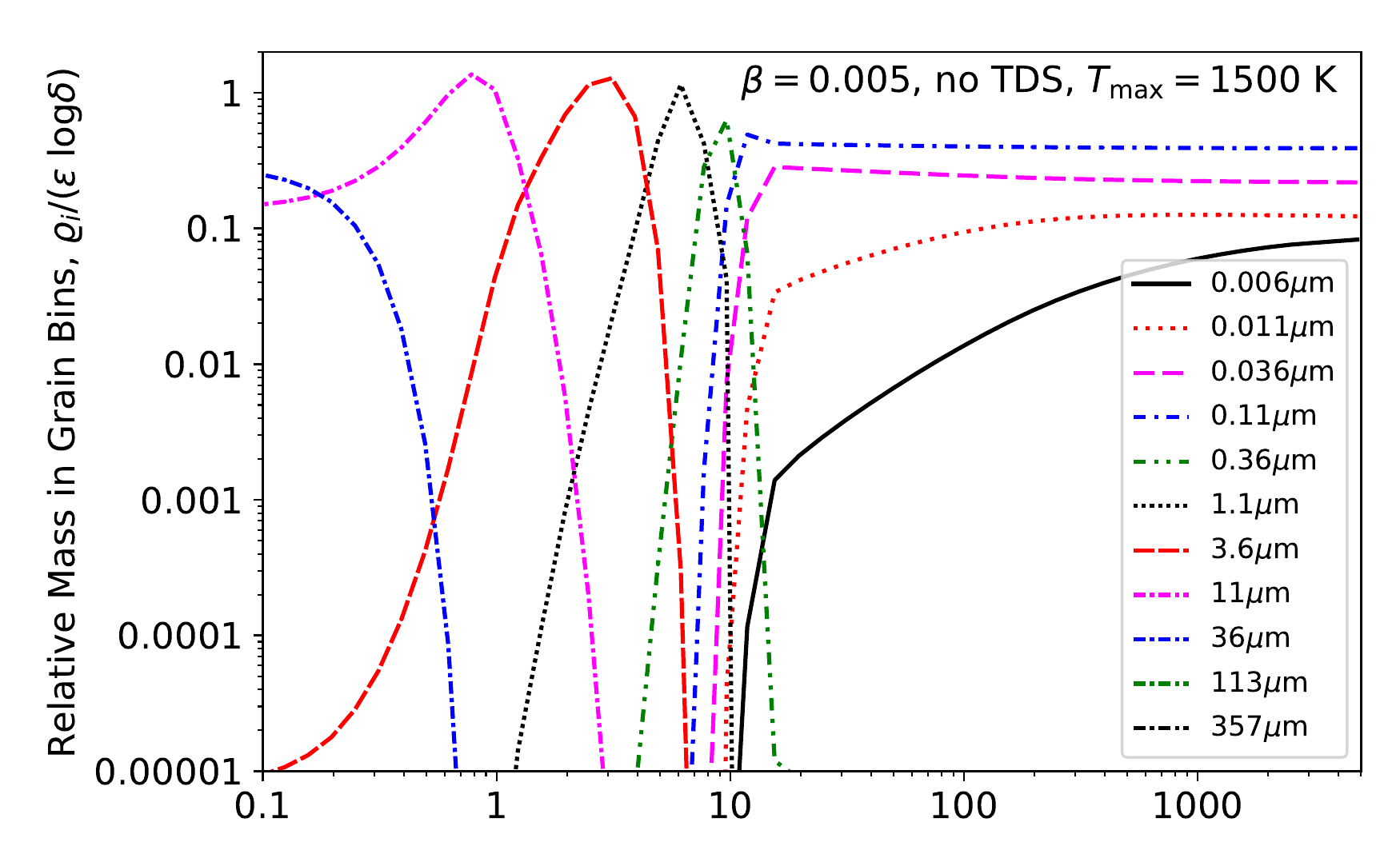}         \vspace{-0.3cm}
    
    \includegraphics[width=7.7cm]{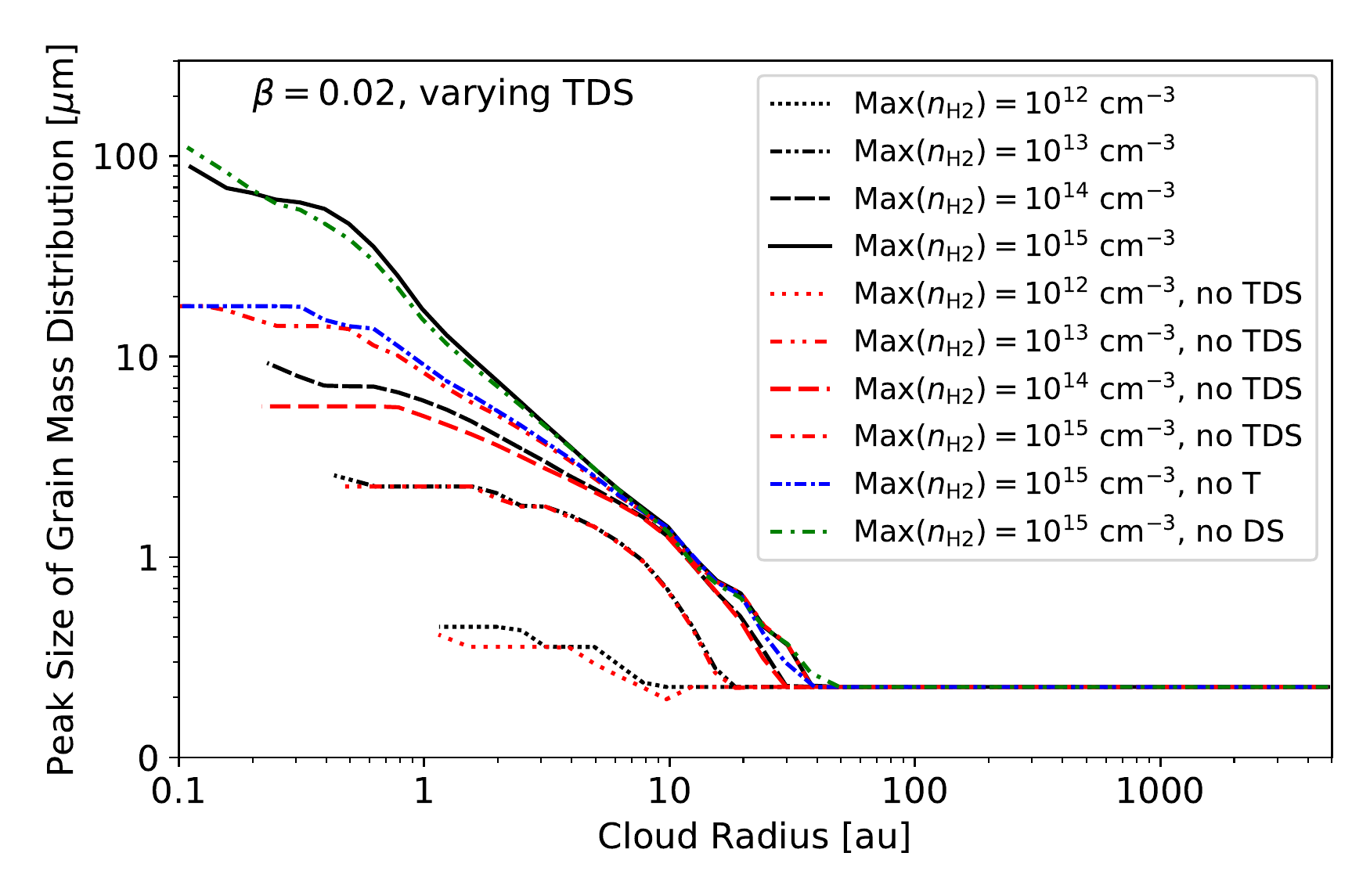}\hspace{0.5cm} 
    \includegraphics[width=7.7cm]{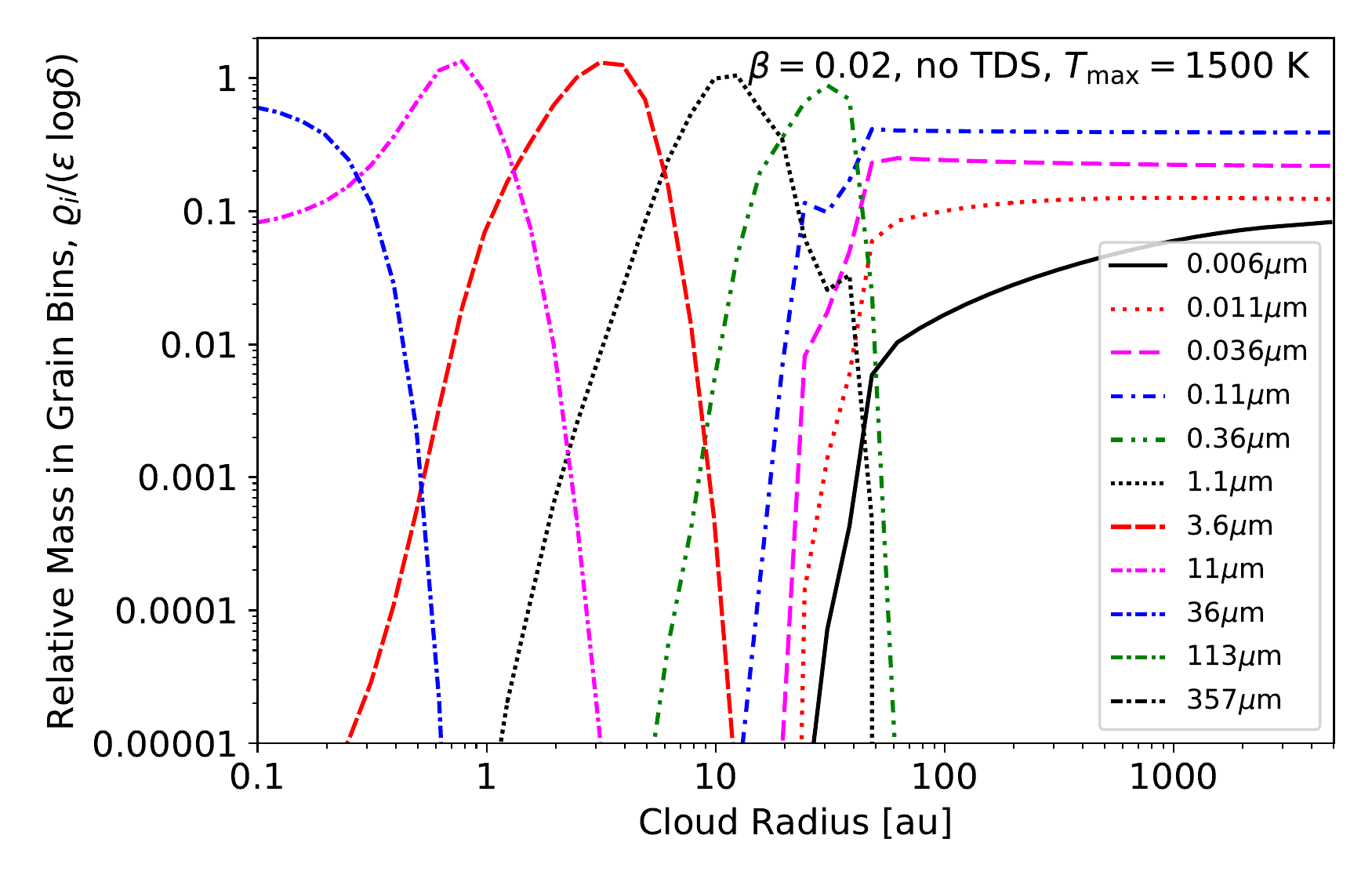}         \vspace{-0.3cm}
    
\caption{The same as Fig.~\ref{fig:rot_mass_bins}, but showing the effects of turning off the turbulence, radial drift, and vertical settling contributions to the grain relative velocities.   Left panels: the peak sizes of the dust grain mass distributions as functions of radius near the mid-plane at four times (denoted by the maximum number density, $n_{\rm H2}$) during the evolution of the rotating molecular cloud cores ($\beta=0.005, 0.02$ from top to bottom) after the formation of the first hydrostatic cores.   The curves denoted `no TDS' are from calculations that exclude turbulent, drift, and settling contributions to the grain velocities.  The curves denoted `no T' omit only the turbulent (envelope and disc) contributions, while the curves labelled `no DS' omit only the radial drift and vertical settling contributions.  Right panels: the relative dust masses in selected grain size bins as functions of radius near the mid-plane  when the maximum temperature has reached 1500~K.  The calculations used to produce these right panels omit turbulent, radial drift and vertical settling contributions to the grain relative velocities (i.e., they are `no TDS' models).   }
\label{fig:TDS}
\end{figure*}

\subsection{Rotating collapse}
\label{sec:rotating}

During the collapse of a non-rotating molecular cloud core, the time allowed for grain growth is minimal -- the time between the maximum density reaching $n_\mathrm{H2} = 10^{11}$~cm$^{-3}$ and the onset of the second collapse phase is only $\approx 1000$~yrs.  In reality, molecular cloud cores have some rotation \citep{Goodmanetal1993} which will lead to molecular cloud cores collapsing to produce rotating first hydrostatic cores and/or pre-stellar discs before the stellar core forms \citep{Bate1998, SaiTom2006, SaiTomMat2008, MacInuMat2010, MacMat2011, Bate2011} and, eventually, protostellar and long-lived protoplanetary discs.

The evolution of the maximum density with time is shown in Fig.~\ref{fig:density_time} for all of the cloud calculations ($\beta=0,0, 0.0025$, 0.005, 0.01, 0.02, 0.05).  The more rapidly rotating clouds take somewhat longer to collapse, as expected due to the extra rotational support.  Molecular hydrogen dissociation above temperatures of $\approx 2000$~K triggers the rapid second collapse phase within the first hydrostatic core that allows the formation of the stellar core \citep{Larson1969}.  The additional time allows dust grains to grow to larger sizes within the first hydrostatic core/pre-stellar disc.  However, it is not just the extra time that matters.  In the non-rotating cloud, the only sources of grain relative motion were due to Brownian motions (that dominate for very small grains; Section \ref{sec:Brownian}), motions induced by gas turbulence (more important for the larger initial grains) and due to the radial pressure gradient within the envelope and first hydrostatic core which leads to terminal infall speeds that depend on grain size (although, as discussed in the previous section, we find that effect of this last source of relative grain velocities is negligible).  However, in the rotating clouds, new sources of dust grain relative velocities become available: pressure gradients in a disc, leading primarily to radial migration and vertical settling (Section \ref{sec:pres_disc}), and disc turbulence (Section \ref{sec:turbulence}).

\subsubsection{Grain size distributions for the rotating models}

The additional sources of relative grain velocities and the extra time available allows for peak grain sizes of up to $\approx 0.5$~mm to be reached in the highest density regions before the temperature reaches 1500~K in some of the rotating cloud calculations.  In Fig.~\ref{fig:rot_mass_bins} we provide plots of the peak grain size as a function of radius at several points during the collapse for each of the rotating clouds (left panels), and we also give the radial distributions of the relative masses of various grain size bins when the maximum temperature reaches 1500~K for each of the rotating clouds (right panels).  These distributions provide the mass-weighted mean quantities within $\pm 5^\circ$ of the midplane (the $xy-$plane).  As in the $\beta=0$ calculation, the peak grain size only begins to increase within the first core or pre-stellar disc, although the smallest dust grains are depleted within the inner few hundreds of au of the collapsing envelope and turbulence-driven coagulation is able to produce a significant population of grains with sizes of $s\approx 0.3-0.5$~$\mu$m within the inner few hundreds of au of the envelope.

\begin{figure*}
\centering 
    \includegraphics[width=16cm]{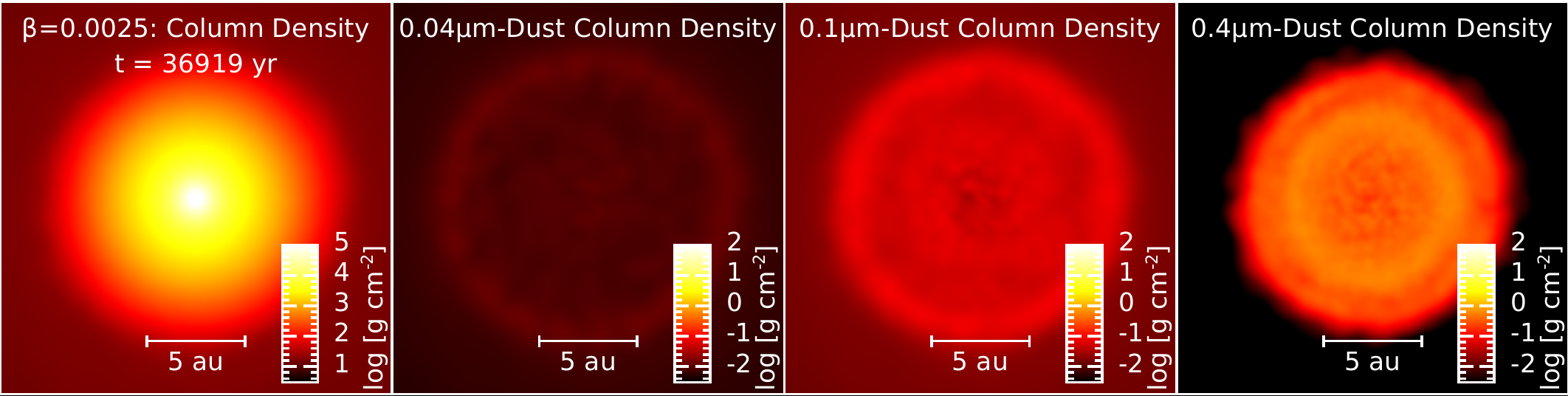}    
    \includegraphics[width=16cm]{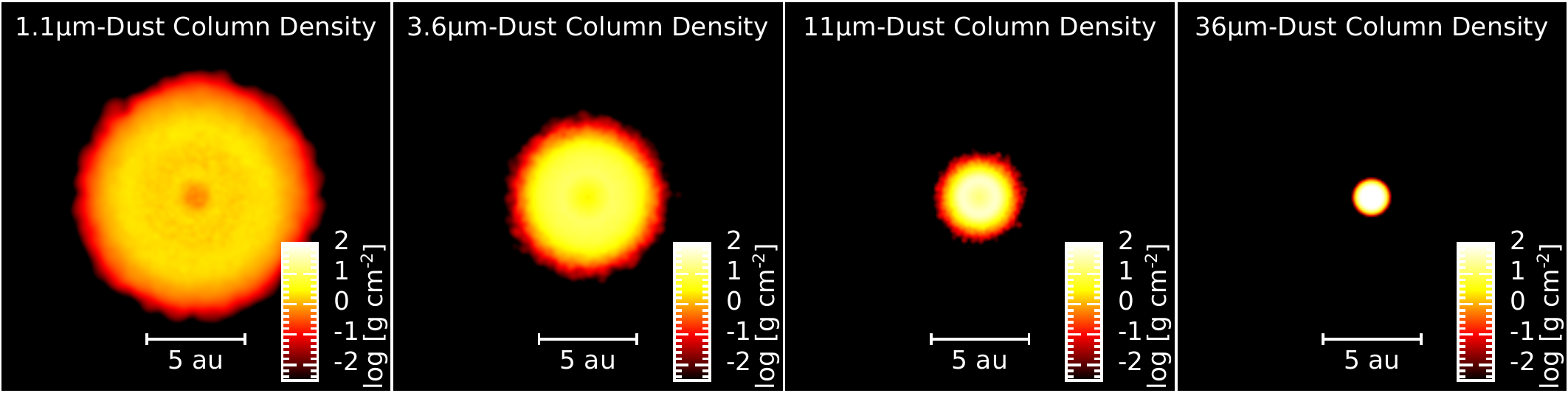}\caption{The logarithm of the column density projected parallel to the rotation axis when the $\beta=0.0025$ molecular cloud core has collapsed to produce a first hydrostatic core with a central temperature of 1500~K (maximum number density $n_\mathrm{H2} \approx 10^{15}$~cm$^{-3}$).  The top left panel gives the total column density (i.e., gas plus dust), while the remaining panels show the column densities of dust grains in selected grain size bins ($s=0.036, 0.11, 0.36, 1.1, 3.6, 11, 36~\mu$m).  The smallest grains are depleted within the first hydrostatic core, while progressively larger grains are found closer to the centre.}
\label{fig:beta0025}
\end{figure*}

Within the first core/pre-stellar disc, the grain populations become mono-disperse with larger grains found at progressively higher densities/smaller radii.  Within the inner 1~au just before the onset of the second collapse, most grains have sizes $s = 10-100\mu$m in the most slowly rotating cloud $\beta=0.0025$.  For the more rapidly-rotating clouds, the typical sizes  within the inner 1~au increase further to $s \approx 40-400~\mu$m for $\beta=0,01, 0.02$.  Even grains with sizes of $s=0.5$~mm are reasonably abundant within the inner au for the intermediate rotation rates ($\beta \geq 0.01$), {\em before the stellar core forms}.  Grains with sizes $s \gtrsim 10~\mu$m develop over a very short amount of time: the time for maximum density in the rotating calculations to go from $n_{\rm H2} = 10^{14}$ to  $10^{15}$~cm$^{-3}$ is between $500 -1000$~yr (Fig.~\ref{fig:density_time}).

Earlier, we found that in the non-rotating collapse the relative grain velocities driving grain growth were dominated by Brownian and turbulence-driven motions.  In the rotating clouds, we include prescriptions for sources of relative grain velocities due to disc turbulence, radial drift, and vertical settling (Sections \ref{sec:turbulence} -- \ref{sec:pres_disc}).  What effects do these sources of relative grain velocities have?  In Fig.~\ref{fig:TDS}, we again plot the peak grain size as a function of radius at several times (left panels) and the radial distributions of the relative masses of various grain size bins, but this time we include results from calculations in which the turbulence (both envelope and disc contributions) and/or radial drift and vertical settling were switched off.  In the interests of brevity, we only plot graphs for rotating clouds with $\beta=0.005, 0.02$.  For these two rotation rates, we present results from calculations that excluded turbulence (`no T'), or excluded radial drift and vertical settling (`no DS'), or that excluded all three contributions (`no TDS').  These results are compared to each other, and to the results that were obtained with all three contributions included that were already presented in Fig.~\ref{fig:rot_mass_bins}.  The three disc contributions to the radial grain velocities make almost no difference to the results until the maximum density exceeds $n_{\rm H2} \approx 10^{14}$~cm$^{-3}$.  In all of the rotating calculations, the peak grain size when the maximum density is $n_{\rm H2} = 10^{14}$~cm$^{-3}$ is $s \approx 5-8~\mu$m (and $s \approx 4~\mu$m with $\beta=0$).   Even beyond this point, it is only within the inner $r \approx 2-3$~au of the first hydrostatic core that significant differences are found.  As expected, more sources of relative grain velocities gives greater dust growth.  Note, however, that it is the addition of the (disc) turbulent velocities that makes most of the difference; the inclusion or omission of radial drift and vertical settling makes very little difference.  Without the contributions from disc turbulence, the grain sizes are restricted to $s \approx 10-40~\mu$m in the centre of the first hydrostatic core, compared to typical grain sizes of $s \approx 40-400~\mu$m when the effects of disc turbulence are included.  Our prescription for the disc turbulence assumes an $\alpha$-disc model with $\alpha_{\rm SS}=0.001$.  Larger or smaller values of $\alpha_{\rm SS}$ would be expected to have more or less of an effect, respectively.  Regardless, as with the non-rotating collapse, at this early time in the star formation process, assuming the molecular cloud core only contains small dust grains initially, most of the grain growth is captured  by considering only Brownian motions.  Turbulence-driven growth plays a small role in the collapsing envelope, and also plays a significant role deep within the first hydrostatic core (or at the centre of a pre-stellar disc) very shortly before the temperatures exceed 1500~K and the second collapse phase begins.

\begin{figure*}
\centering 
    \includegraphics[width=16cm]{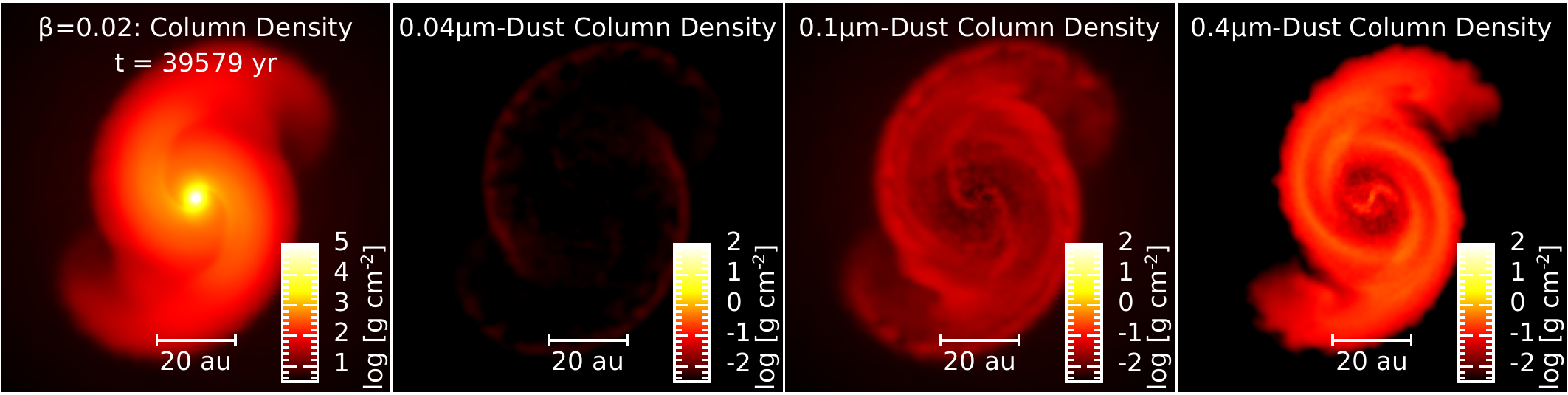}    	
    \includegraphics[width=16cm]{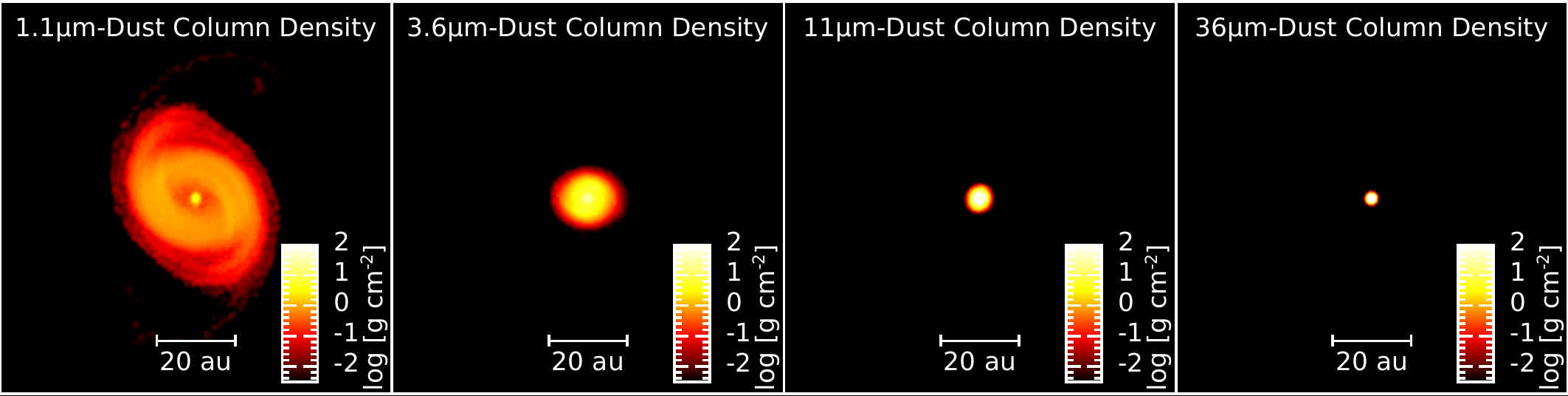}
\caption{The logarithm of the column density projected parallel to the rotation axis when the $\beta=0.02$ molecular cloud core has collapsed to produce a gravitationally-unstable pre-stellar with a central temperature of 1500~K (maximum number density $n_\mathrm{H2} \approx 10^{15}$~cm$^{-3}$).  The top left panel gives the total column density (i.e., gas plus dust), while the remaining panels show the column densities of dust grains in selected grain size bins ($s=0.036, 0.11, 0.36, 1.1, 3.6, 11, 36~\mu$m). The smallest grains are absent from the pre-stellar disc, intermediate sized grains are found in the spiral arms and outer disc, and the largest grains are found in the central core.}
\label{fig:beta02}
\end{figure*}

\begin{figure*}
\centering 
    \includegraphics[width=16cm]{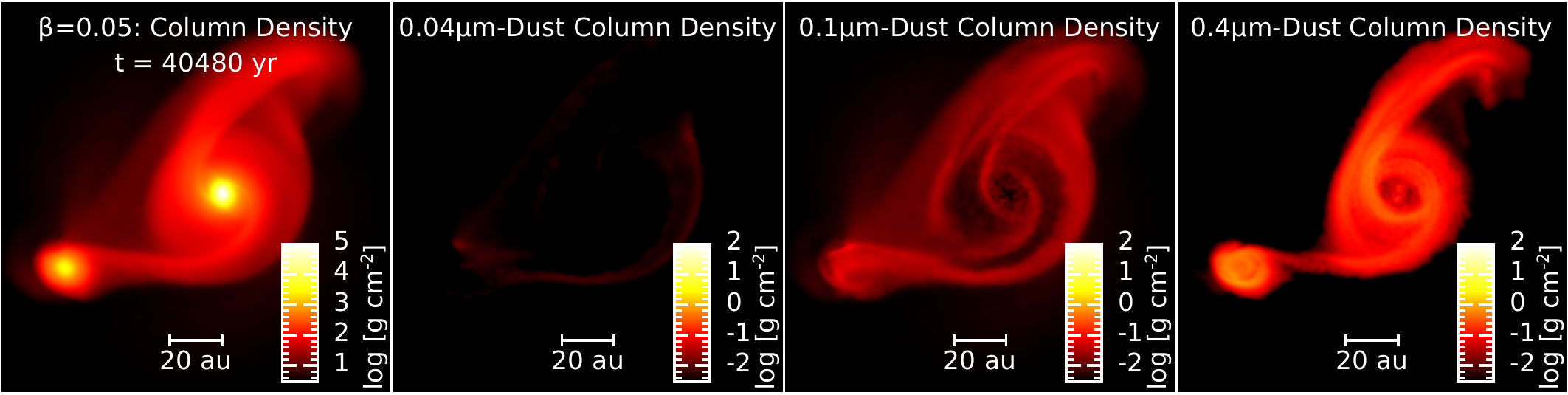}    	
    \includegraphics[width=16cm]{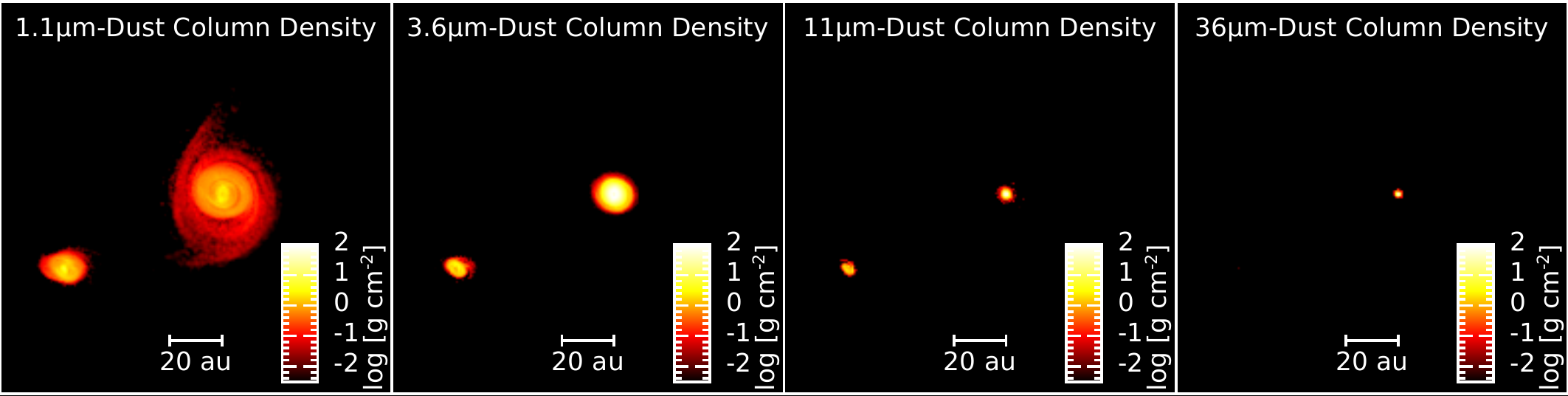}
\caption{The logarithm of the column density projected parallel to the rotation axis when the $\beta=0.05$ molecular cloud core has collapsed to produce a first hydrostatic core with a maximum number density $n_\mathrm{H2} \approx 10^{14}$~cm$^{-3}$.  The top left panel gives the total column density (i.e., gas plus dust), while the remaining panels show the column densities of dust grains in selected grain size bins ($s=0.036, 0.11, 0.36, 1.1, 3.6, 11, 36~\mu$m). The smallest grains are absent from the pre-stellar disc.  Within the spiral arms, smaller grains are found at the edges of the arms with larger grains deep within the arms, while the largest grains are found in the central core.  Some rings of intermediate-sized ($s \approx 1~\mu$m) can be seen.  One of the arms is beginning to fragment to form a companion protostar, and rapid grain growth is also occurring within it.}
\label{fig:beta05}
\end{figure*}

\subsubsection{Projected grain size distributions}
\label{sec:projected}

In Figs.~\ref{fig:beta0025}--\ref{fig:beta05}  we show images of the column density as viewed parallel to the rotation axis (looking down on the discs).  We only provide figures for the $\beta=0.0025, 0.02, 0.05$ cases.  The $\beta=0.005, 0.01$ cases look similar to the $\beta=0.0025$ case except that the pre-stellar discs have somewhat larger radii and there is a very slight sign of gravitational instability at their outer edges (i.e., the outer regions develop weak non-axisymmetric structure).  In each set of 8 panels, the first panel gives the total column density (i.e., essentially the gas column density), while the other panels give the column densities of the dust grains contained in various grain size bins, with grain size increasing from left to right, and down.  Each of the plots is made when the maximum temperature reaches 1500 K, except for the $\beta=0.05$ case which we show when its maximum number density reaches $n_{\rm H2}=10^{14}$~cm$^{-3}$ due to the added complexity of the disc fragmentation that occurs in this calculation.  It is clearly seen that the largest dust grains are located closer to the centres of the first hydrostatic cores.  This has nothing to do with the radial migration of dust grains that would be expected for larger grains in protoplanetary discs.  Instead, this is simply the result of the strong dependence of grain growth on the number density of the dust grains; dust grows faster in higher density regions. Conversely, very small grains with $s=0.04, 0.1$~$\mu$m, and even $s=0.4$~$\mu$m with the higher rotation rates, are substantially depleted within the first hydrostatic cores or the central regions of the pre-stellar discs.

For the more rapidly-rotating molecular cloud cores ($\beta=0.02, 0.05$), the pre-stellar discs become dynamically rotationally unstable due to their high ratio of rotational to gravitational potential energy \citep[e.g.,][]{Bate1998,Bate2011}.  In these cases, an axisymmetric pre-stellar disc forms initially and then deforms into a bar-shaped object whose outer parts form trailing spiral arms.  Gravitational torques from the spiral arms remove angular momentum from the inner parts of the object and the result is a dense central core surrounded by a large disc with strong spiral arms.  This can be seen in Figs.~\ref{fig:beta02}, \ref{fig:beta05}.  At any given radius from the central object the larger grains are preferentially located within the dense spiral arms.   Again, this has nothing to do with the migration of dust grains toward pressure-maxima.  The grains are too small to undergo significant migration.  Instead, the higher grain densities within the spiral arms is enough to result in substantially faster dust growth within the spiral arms.  For example, in the most rapidly-rotating case ($\beta=0.05$), at radii $r \gtrsim 20$~au from the centre of the pre-stellar disc, the grains with sizes $s \approx 0.4$~$\mu$m are confined to the dense spiral arms.  Note that the left spiral arm is undergoing fragmentation to form a companion protostar, and it too is producing a larger grains (up to $s\approx 10$~$\mu$m at this early stage).  

Conversely, the faster dust growth within the arms can also produce a deficit of small grains within the dense arms.  Again, this is particularly apparent in the most rapidly-rotating case ($\beta=0.05$) where the $s=0.1$~$\mu$m dust grains are abundant at the edges of the arms, but strongly depleted within the arms themselves.  

Finally, in the most rapidly-rotating case we even find rings and arcs of $s\approx 1$~$\mu$m dust grains forming at radii of $r=10-20$ au (the lower left panels of Fig.~\ref{fig:beta05}).  These rings and arcs are produced by the winding up of earlier spiral arms and of the comparatively large grains that were produced within them.  

As online supplementary material to accompany this paper, we provide animations of the evolution of all five rotating calculations.  The animations show the same 8 panels as in Figs.~\ref{fig:beta0025}--\ref{fig:beta05}.  The animations are useful for seeing: how the grains grow first at small radii (high densities); the development of the bar and spiral arm instabilities; and how rings of intermediately-sized dust grains form from the winding up of spiral arms in both the $\beta=0.02$ and $\beta=0.05$ calculations.

In summary, we find that dust growth can be very rapid inside the first hydrostatic core/pre-stellar disc.  These protostars are so young that they haven't even formed a stellar core yet.  The grain growth is driven by the Brownian and turbulence-driven motions of small dust grains. The larger dust grains are found in the higher density regions, meaning that in general the grains are larger closer to the centre of the core/disc.  But if the discs contain strong spiral density waves, rapid grain growth within the spiral arms can produce a very diverse spatial distributions of grain sizes.

\subsection{Caveats}
\label{sec:caveats}

The calculations performed for this paper do not include magnetic fields.  The primary role of magnetic fields in terms of the dynamics of the collapse is to remove angular momentum from the central regions (magnetic braking) and to drive outflows.  For calculations similar to those presented here, studying the first hydrostatic core phase, the angular momentum transport typically manifests as first hydrostatic cores that are shorter-lived, less disc-like, and less likely to be gravitationally unstable for a given initial cloud rotation rate \citep*[e.g.,][]{BatTriPri2014, WurBatPri2018a, WurBat2019}, although for calculations that include the Hall effect whether the magnetic field inhibits or enhances disc formation depends on whether the magnetic field is aligned or anti-aligned with the rotation axis, respectively \citep{WurBatPri2018a}.  In terms of the outflows, slow ($\sim 1-2 $~km~s$^{-1}$) bipolar outflows are found to be launched during the first hydrostatic core phase  (\citealt{Tomisaka2002, Machidaetal2005, BanPud2006}; \citealt*{MacInuMat2006, MacInuMat2008}; \citealt{HenFro2008, Commerconetal2010, Burzleetal2011b, Tomida_etal2013, BatTriPri2014, WurBatPri2018a}).  Despite these additional dynamical effects, the conclusions obtained by examining the purely hydrodynamical calculations presented here should be qualitatively unchanged, unless the inclusion of magnetic fields strongly affects the distributions of relative grain velocities (c.f., \citealt{Guillet_etal2020}).  One aspect that it would be interesting to study in the future is whether large dust grains can be launched from the first core/disc in magnetically-driven outflows as suggested by \citep{WonHirLi2016}.  In this way, comparatively large dust grains (e.g., tens of microns to millimetre sizes) could be launched above and below the disc, into the envelope, and subsequently fall back onto the outer disc to `seed' a large-grain population.  We leave this possibility to a later study.

The grain model we use in this paper assumes spherical grains with a single composition (we use intrinsic density of 2.26~g~cm$^{-3}$, appropriate for graphite grains).  In reality, small dust grains are likely to have complicated geometries, often described as having fractal structure \citep*{KemPfaHen1999}.  This means that for a given mass, grains will have a larger cross section.  This will have two main affects on the grain growth.  First, larger cross sections mean that grain growth through collisions can occur more easily.  Second, however, larger cross sections also mean that the grains will be more tightly coupled to the gas, potentially leading to smaller relative velocities.  But the dust-gas coupling (i.e., the stopping time or Stokes number) is important for determining the relative motions due to gas drag such as terminal velocities, turbulent velocities, radial drift and vertical settling.  Since we have shown that grain growth at these early times (given the initial small grain sizes) is driven by Brownian motion (which shouldn't be affected in the same way), it is likely that the growth of fractal grains to sizes of $s \sim 10~\mu$m would occur even more rapidly than we have found here due to the larger cross sections.  However, again we believe that the qualitative conclusions would be unchanged. 

Similarly, although in reality dust grains will differ in composition and will have different sticking and strength parameters \citep[e.g.,][]{ChoTieHol1993}, this should not affect our qualitative conclusions.  Because we only need to consider small grains that have relatively low velocity dispersions, we have assumed perfect sticking and the strength of the grains is irrelevant because they do not suffer collisions that are fast enough to fragment them.  Fragmentation would need to be considered if the calculations were evolved to the point that large grains were present at lower densities (i.e., into the protoplanetary disc regime).  However, before that regime was reached, our assumption that dust grains do not migrate substantially compared to the spatial resolution of the calculations would break down and grain migration would also need to be modelled.

\section{Conclusions}
\label{sec:conclusions}

For the first time, we have performed three-dimensional radiation hydrodynamical calculations of the collapse of molecular cloud cores that include a treatment to model dust grain growth.  Our method treats the grain growth in each parcel of gas independently, and is valid so long as the dust grains don't grow large enough to migrate over the spatial scales resolved by the simulation during the timescales that are followed.  Beginning with a typical MRN dust grain size distribution, ranging from 5~nm to 0.25~$\mu$m with most of the mass in the largest grains, we examine the spatial and temporal growth of the grains during the collapse of 6 molecular cloud cores with different rotation rates to form first hydrostatic cores or pre-stellar discs.  We only analyse the calculations up to the point where the maximum temperature reaches 1500~K (since above this temperature dust would be expected to evaporate).  Our main conclusions are:

\begin{enumerate}
\item We find that the peak grain size (in terms of mass) does not grow significantly within the collapsing envelopes (i.e., it remains at $s\approx 0.25~\mu$m).  However, small grains become depleted as the gas collapses, with the smallest grains decreasing in abundance by one to two orders of magnitude within the inner 100 au.
\item Large dust grains only begin to grow significantly within the first hydrostatic core or pre-stellar disc, but they do so very quickly.  The high densities within the first core/disc allow substantial growth on short timescales of only hundreds or thousands of years.
\item Within the first hydrostatic core or pre-stellar disc, the grains evolve to a locally-monodisperse (peaked) distribution.  Small dust grains are eliminated, and the typical grain size generally increases with decreasing radius from the centre of the core.  Within the inner 1 au, the dust grain sizes range from $s \approx 10-100~\mu$m or even $s \approx 40 -400~\mu$m, with substantial abundances of grains with sizes up to $s \approx 0.5$~mm in some of the models (those with larger initial rotation rates).
\item Due to the small initial grain sizes, grain growth is driven almost exclusively by relative grain velocities due to Brownian motion until sizes of $s \approx 10~\mu$m, although turbulence-driven coagulation does allow a population of grains with sizes up to $s\approx 0.5~\mu$m to develop in the inner parts of the collapsing envelope.  Turbulent stirring of the grains also plays a role deep within the first core (the inner few au) and is necessary for substantial populations of $s\gtrsim 100~\mu$m grains to be formed prior to the onset of the second collapse phase.  Other sources of relative grain velocities such as those due to terminal infall in the envelope, and radial and azimuthal drift and vertical settling in a disc are insignificant at these early times.
\item Due to the relatively small grain sizes and/or high gas densities, grain motions relative to the gas (such as those from terminal infall velocities, disc turbulence, and radial/azimuthal migration and vertical settling in young discs) remain low enough throughout these early stages of star formation that any migration of grains relative to the gas (or each other) is insignificant compared to the spatial resolution of the calculations (Appendix \ref{appendixB}).  Thus, we are justified in our use of a method that omits dust migration.
\item The preference for larger grains at smaller radii from the centre of the first core/pre-stellar disc is not due to radial migration (which is negligible at these stages), it is simply due to differential grain growth due to the radial density and temperature gradients.
\item In rapidly-rotating first hydrostatic cores that are disc-like (i.e., pre-stellar discs) and are gravitationally unstable, grain growth is more rapid in the spiral density waves leading to larger grains being preferentially found in the spiral arms even though there is no migration of grains relative to the gas.  Thus, the grain size distribution can vary substantially in the disc even at these very young ages before the star itself has formed, and this variation is not due to grain migration into regions with pressure maxima.
\end{enumerate}

Finally, we speculate that if magnetic fields were included, the outflows they launch from the first hydrostatic core may potentially entrain and launch some of the large dust grains that form in the first core/pre-stellar disc, up into the envelope where they could fall back onto the disc at large distances and potentially provide a source of large grains at large radii earlier than if dust grains had to slowly grow through in situ coagulation.  If some of these dust grains came from the hot inner region of the first core, they may even provide a source of some of the chondrules that are found in Solar System meteorites.

\section*{Acknowledgments}

MRB thanks Sebastiaan Krijt, Adrien Houge and Pablo Lor\'en-Aguilar for useful conversations, and James Wurster for  comments that helped improve the manuscript.  MRB also acknowledges the input of the anonymous referee who suggested including the effects of turbulence in the envelope.

The rendering of the calculations were produced using SPLASH \citep{Price2007}, an SPH visualization tool publicly available at http://users.monash.edu.au/$\sim$dprice/splash.

This work was supported by the European Research Council under the European Community's Seventh Framework Programme (FP7/2007-2013 Grant Agreement No. 339248).  This work used the University of Exeter Supercomputer, Isca.  This research was supported in part by the National Science Foundation under Grant No. NSF PHY-1748958.

For the purpose of open access, the author has applied a Creative Commons Attribution (CC BY) licence to any Author Accepted Manuscript version arising.

\section*{Data Availability}

The data used to produce Figs.~\ref{fig:tvalid}--\ref{fig:TDS} and Figs.~\ref{A1} are provided as Additional Supporting Information (see below).  The SPH data files that are required to produce Figs.~\ref{fig:beta0025}--\ref{fig:beta05} and  Fig.~\ref{B1} are available from \cite{Bate2022_data}.

\bibliography{mbate}

\section*{Supporting Information}

Additional Supporting Information may be found in the online version of this article: \\
{\bf Animations.} We provide animations of the gas and dust column densities projected parallel to the rotation axis for all five of the rotating calculations.  These animations show the evolution with time of the same quantities as shown in Figs.~\ref{fig:beta0025}--\ref{fig:beta05}, ending with the views provided in the figures. \\
{\bf Data files and plotting routines.} We provide the data text files and Python routines used to produce Figs.~\ref{fig:tvalid}--\ref{fig:TDS} and Fig.~\ref{A1}.  We also provide the SPLASH configuration files used to render  Figs.~\ref{fig:beta0025}--\ref{fig:beta05} and the animations, and to produce Fig.~\ref{B1}.

\appendix

\section{}
\label{appendixA}

\begin{figure}
\centering \vspace{-0.25cm} \vspace{0cm}
    \includegraphics[width=8.8cm]{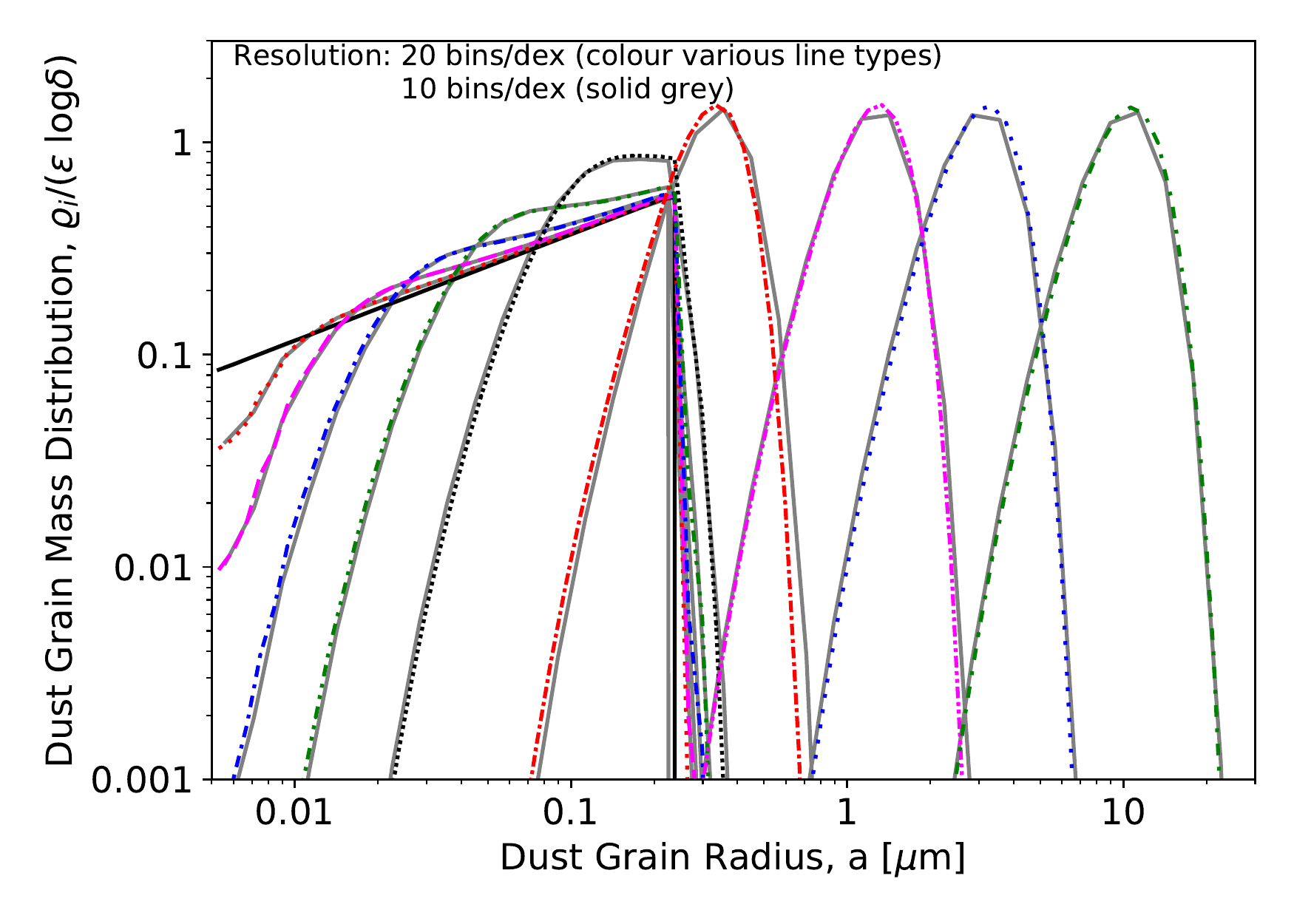}         \vspace{-0.5cm}
\caption{The evolution of the dust grain mass distribution function at the centre of the spherically collapsing (non-rotating, $\beta=0$) molecular cloud core as a function of the maximum molecular hydrogen number density, $n_\mathrm{H2}$, during the collapse.  We show the results from two calculations:  a calculation performed using 86 dust bins ranging from 5~nm to 100~$\mu$m (i.e., 20 bins per decade of dust size; same data as Fig.~\ref{fig:massdist_beta0}), and a calculation using 53 dust bins ranging from 5~nm to 1~mm (i.e., 10 bins per decade of dust size; grey solid lines).  The rate of growth and the size distributions are well modelled with either resolution. The legend has been omitted for clarity, but for the non-grey lines it is the same as in Fig.~\ref{fig:massdist_beta0}.}
\label{A1}
\end{figure}

In this paper, the dust distribution associated with each SPH particle is modelled using logarithmically-spaced bins.  A question that should be asked is: how sensitive is the dust growth to the number of bins?  Using more bins provides better resolution of the grain size distribution, but it also requires more memory and computational time.  

In Fig.~\ref{A1}, we plot the grain size distribution at the centre of the non-rotating ($\beta=0$) cloud each time the maximum (central) density has increased by an order of magnitude.  These calculations exclude the contribution of turbulence-driven relative grain velocities. We give the results obtained using two different simulations: a low-resolution simulation where the cloud is modelled using $10^6$ SPH particles and the dust grain distribution is modelled using 10 bins per decade of grain radius (53 bins ranging from 5~nm to 1~mm); and a high-resolution simulation where the cloud is modelled using $3 \times 10^6$ SPH particles and the dust grain distribution is modelled using 20 bins per decade of grain radius (86 bins ranging from 5~nm to 100~$\mu$m).  The latter results are the same results as those plotted in Fig.~\ref{fig:massdist_beta0}.  It can be seen that although the grain size distribution is not as smoothly modelled with fewer bins, there is no systematic change in the distributions and the same dust growth rate is obtained in both simulations.  

Therefore, in the bulk of the simulations presented in this paper we employ $3 \times 10^6$ SPH particles so that the gas radiation hydrodynamics is well modelled, but we use only 53 bins ranging from 5~nm to 1~mm (10 bins per decade) to model the dust size distribution.  As shown by Fig.~\ref{A1}, this is adequate to capture the grain growth, but it requires substantially less memory and somewhat less computational time that would be required for 20 bins per decade.

\section{}
\label{appendixB}

In Section \ref{sec:limitations}, we argue analytically that we can treat the dust grain populations associated with each SPH particle as evolving as independent `zones' and neglect the migration of dust grains between resolution elements over the timescales we model.  However, we don't just rely on this analytic argument -- we check throughout the simulations that the dust grains would not have migrated a substantial fraction of a resolution length.

The resolution length of an SPH simulation is approximately equal to the SPH particle smoothing length, $h$, which varies in space and time (SPH is a Lagrangian method).  Therefore, during the simulation for each SPH particle we monitor how far the typical dust grain in that element of gas may have moved as a fraction of the smoothing length to confirm that this is less than the resolution length.  We record the distance as a fraction of the smoothing length due to the Lagrangian nature of the SPH particles, in particular that the smoothing length becomes smaller as density increases within the collapsing cloud.

In practice, for each SPH particle, $j$, we integrate
\begin{equation}
\frac{{\rm d}X_j}{{\rm d}t} = \frac{v_{{pm},j}}{h_j},
\end{equation}
from the beginning of the simulation, $t=t_0$, to the end of the simulation, $t=t_{\rm end}$, so that at the end of each simulation for each SPH particle we have 
\begin{equation}
X_j = \int_{t_0}^{t_{\rm end}} \frac{v_{{pm},j}}{h_j} ~{\rm d}t.
\end{equation}
$X_j$ is then the distance that a typical dust grain may be expected to have migrated as a fraction of the smoothing length, $h_j$, of the $j$th SPH particle.  

\begin{figure}
\centering 
     \includegraphics[width=8.5cm]{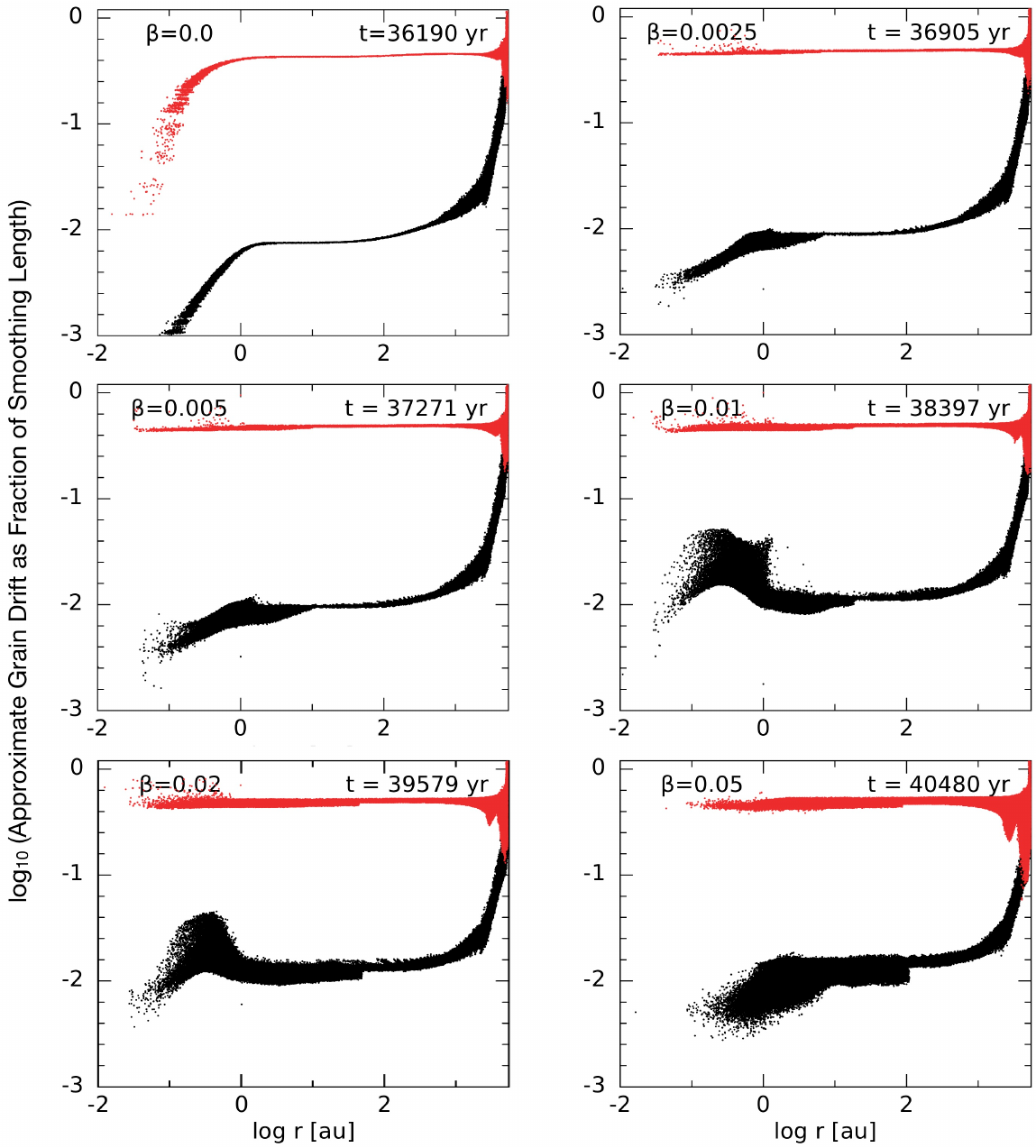}    	
\caption{Estimates of how far (as a fraction of the local resolution length) the typical dust grain may have migrated relative to the gas for each SPH particle during each of the $3\times 10^6$ SPH-particle calculations.  The red (upper) and black (lower) points include or exclude, respectively, grain speeds due to envelope turbulence.  Except for the outer parts of the clouds (where grain growth is insignificant) or turbulence-induced motions very early in the collapse, before the grain population evolves significantly, any dust grain migration should be orders of magnitude smaller than the local resolution length. This validates our approach of treating the dust growth associated with each parcel of gas independently. }
\label{B1}
\end{figure}

The quantity $v_{{pm},j}$ is the velocity difference between a dust grain at the peak of the dust mass distribution for SPH particle, $j$, and the minimum possible dust grain size ($\approx 5$~nm in the calculations performed for this paper).  This is calculated using the same equation \ref{eq:quadrature} that we use to calculate relative grain velocities, except that we omit the Brownian motion term.  Since the smallest grains will be well coupled to the gas, this essentially provides a measure of the velocity difference between the gas and a grain with the typical size in that `zone' (i.e., at the peak of the grain mass distribution).  The Brownian motion term must be omitted because the Brownian motion velocities are largest for the smallest particles.  For all other sources of velocity difference between the dust and the gas (i.e., terminal infall velocities, turbulence, radial drift and vertical settling) the smallest grains follow the gas, so computing $v_{{pm},j}$ without the Brownian motion term gives a measure of the relative velocity between the typical dust particle and the gas.  Moreover, the relative velocity due to any source other than Brownian motion between any dust grain that has a mass in between the smallest grain and the typical grain size should be smaller than this value (for example, for infall at the terminal velocity, the largest grains have the highest speeds).  For Brownian motion itself, because it is generated locally by grains being randomly bombarded by gas particles the random walk is unlikely to result in dust grains migrating significantly compared to the resolution length on the timescales we model.

In Fig.~\ref{B1} we plot the values of $X_j$ for every SPH particle at the end of each of the $3\times 10^6$ SPH-particle calculations with six different rotation rates.  We plot the values that are obtained when excluding the turbulence of the envelope (black points, which tend to have low values) and the values that are obtained when including envelope turbulence (red points).  The latter usually have values $X_j \approx 0.5$.  These high values are due to the evolution at the beginning of the collapse of the cloud -- at low densities the largest grains have comparatively high turbulence-driven velocities since ${\rm St} \propto \rho_{\rm G}^{-1/2}$ (see Section \ref{sec:turbulence}).  Thus, early on, the largest dust grains could potentially move relative to the gas.  However, this is unimportant for three reasons.  First, these motions are turbulent rather than having a consistent direction so although their speed integrated over time is reasonably large the actual distance moved will be more like a random walk rather than linear motion.  Second, early on there is no spatial variation in the dust population so turbulence dust diffusion will not result in any change in the local dust population.  Third,  $X_j < 1$ so even linear motions of this distance are still smaller than the resolution length.

A more useful measure of the expected migration of dust grains relative to the numerical resolution length is given by the black points in Fig.~\ref{B1}.  These values include dust grain speeds due to terminal infall velocities, disc turbulence, radial drift, and vertical settling.
In all of the calculations, the values of $X_j \approx 1$ near the outer radius of the molecular cloud core show that here the largest dust grains would be expected to migrate radially relative to the gas (due to infall at their terminal velocity through the low-density gas) over approximately one smoothing length during the calculations.  However, since the grain growth is insignificant in the outer parts of the clouds this is unimportant.  For all except the outer parts of the clouds ($r \lesssim 4000$~au), the typical dust grains whose growth is being modelled are expected to migrate only a small fraction of the resolution length (typically of order 1\%) relative to the gas due to the above sources of grain motions and, therefore, even less distance relative to smaller dust grains.  Thus, we are justified in our assumption that dust grains do not migrate significantly within the simulations over the timescales that we model.

Note that in the $\beta=0.01, 0.02, 0.05$ calculations, there is an increase in the $X_j$ values given by the black points (i.e., excluding envelope turbulence) within a few au of the centre.  There are also somewhat larger values with greater dispersions within $r \approx 20$~au for $\beta=0.01$, $r \approx 50$~au for $\beta=0.02$, and $r \approx 100$~au for $\beta=0.05$.  These regions are where the grain velocity contributions due to disc turbulence, radial drift and vertical settling become significant towards the end of these more rapidly-rotating calculations. Eventually, the largest grains in the first hydrostatic cores or discs would migrate over length scales greater than the numerical resolution length.

\end{document}